\documentclass[aps,prb,reprint]{revtex4-1}
\usepackage{amsmath}
\usepackage{graphicx}
\usepackage{float}
\usepackage{physics}
\newcommand{\hl}{\mathcal{H}}
\newcommand{\mz}{\mathcal{Z}}

\newcommand{\leg}{\langle}
\newcommand{\rg}{\rangle}
\newcommand{\up}{\uparrow}
\newcommand{\down}{\downarrow}
\newcommand{\gup}{{\cal G}_{\up}}
\newcommand{\gdn}{{\cal G}_{\down}}
\newcommand{\nn}{\nonumber}
\newcommand{\mb}[1]{\mathbf{#1}}
\newcommand{\bs}[1]{\boldsymbol{#1}}
\newcommand{\ab}{\mathbf{a}}
\newcommand{\cb}{\mathbf{c}}
\newcommand{\db}{\mathbf{d}}
\newcommand{\sbt}{\substack}

\begin{document}
\date{\today}

\author{Amin Ahmadi}

\affiliation{Department of Physics, University of Central Florida,
  Orlando, FL 32816, USA}

\author{Eduardo R. Mucciolo}

\affiliation{Department of Physics, University of Central Florida,
  Orlando, FL 32816, USA}

\title{Microscopic formulation of dynamical spin injection in
  ferromagnetic-nonmagnetic heterostructures}

\begin{abstract}
We develop a microscopic formulation of dynamical spin injection in
heterostructure comprising nonmagnetic metals in contact with
ferromagnets. The spin pumping current is expressed in terms of
Green's functions of the nonmagnetic metal attached to the ferromagnet
where a precessing magnetization is induced. The formulation allows
for the inclusion of spin-orbit coupling and disorder. The Green's
functions involved in the expression for the current are expressed in
real-space lattice coordinates and can thus be efficiently computed
using recursive methods.
\end{abstract}

\maketitle


\section{Introduction}

One of the key elements in any implementation of spintronics is an
efficient source of spin current.\cite{flatte-spintronic-Challenge}
Among the different methods available, dynamical spin injection from a
ferromagnet metal (FM) into an adjacent nonmagnetic metal (NM) has
been theoretically proposed \cite{Brataas_Spin_battery} and
experimentally observed.\cite{mizukami_exp_pump, saitoh_pump,
  Azvedo_SP_experimental, Heinrich_SP_experimental, mosendz_pump} In
this method, in addition to a longitudinal static magnetic field, an
oscillating transverse magnetic field is applied, inducing a
magnetization precession in the FM. Most of the angular momentum
transferred to the FM by the oscillating field is dissipated through
spin-relaxation processes in the bulk, but a small part survives as a
spin current injected into the NM.

The exotic electronic properties of graphene have captured the
attentions of the physics community since the first experiments with
this material.\cite{geim,deheer} High mobility and a long
spin-relaxation length are features that make graphene a promising
passive element for spintronics.\cite{yang-Gr-relaxation} In addition,
the enhancement of spin-scattering processes in graphene by adatoms or
defects,\cite{Hernando_spinRelaxationGr} which yields spin Hall
\cite{Kane_SHE_Gr} and the inverse spin Hall effects, has led to
proposals of graphene-based spin-pumping
transistors.\cite{Ferreira_Gr_spinPump_transistor,semenov-Gr-transistor}

Recent experimental studies \cite{singh_spin_pumping_extended_Gr,
  Singh_Spumping_fitInterface} show an increase in the damping of the
ferromagnetic resonance (FMR) when a graphene sheet is placed in
contact with a FM subject to an oscillating magnetic field. One
interpretation of phenomenon is that part of the precessing
magnetization leaks into the graphene sheet as a spin current,
effectively leading to another channel of magnetization damping in
addition to the relaxation mechanisms existing in the bulk of the FM.


A time-dependent scattering theory \cite{brataas_1,
  Brataas_Spin_battery} based on the general theory of adiabatic
quantum pumping \cite{Buttiker_quantum_pump} relate the increase in
the FMR damping to the magnitude of a phenomenological mixing
conductance parameter; further effort is necessary to describe
microscopically the process of spin pumping into two-dimensional (2D)
materials, as well as to properly quantify the spin current in terms
of materials and interface parameters. A recent study
\cite{Bauer_2Dpumping} applied the time-dependent scattering theory to
spin pumping in a insulating ferromagnet laid on top of a 2D
metal. While insightful, this approach is not suitable for including
disorder and spatial inhomogeneities such as adatoms; and when applied
to graphene, it was confined to the vicinity of the neutrality point.

In this paper we develop a microscopical formulation of spin pumping
from a FM into a NM material. Both the atomic structure of the
materials and the particular geometry of the system can be taken into
account exactly in this formulation. The spin current expression is
written in terms of the Green's function of the NM portion, allowing
one to apply efficient recursive numerical methods for the computation
of spin currents.\cite{eduardoRGF} Another advantage of the
formulation we present is the possibility to include accurate,
microscopic models of spin-orbit coupling in the NM portion, as it
relies on a spatial tight-binding representation of the system.

Another aspect that can be addressed with this formulation is the
distinction between the angular momentum that relaxes at the interface
and the part that flows into the NM. As it was shown in the experiment
by Singh and coauthors,\cite{Singh_Spumping_fitInterface} where a FM
was laid on top of a graphene sheet, even without graphene protruding
away from the FM (when no spin current injection is possible), the
enhancement of damping is significant. This enhancement has been
associated with two-magnnon scattering at the
interface.\cite{hurben_2_magnnon} However, in systems where graphene
protrudes away from the FM, an extra damping has been measured due to
the flow spin current into graphene. An atomistic study of such
phenomenon is needed to discriminate the contribution of spin current
from the surface relaxation in the enhanced damping.

This paper is organized as follows. In Sec. \ref{sec:model}, we use a
one-dimensional tight-binding chain coupled to a magnetic site to
introduce the time-dependent boundary condition problem and to derive
an expression for the spin current based on an equation-of-motion
formulation. The definition of charge and spin currents appropriate to
the problem in hand are discussed in
Sec. \ref{sec:charge-spin-current}. We apply the formulation to a
zero-length system in Sec. \ref{sec:zero_chain} and a finite-length
chain in Sec. \ref{sec:finite-chain}. In Sec. \ref{sec:2D-formalism}
the general expression for the spin current in the 2D system, including
spin-orbit mechanisms is derived. In Sec. \ref{sec:summary-discussion}
we summarize the results and point to future work. Details of the
formulation and some derivations are presented in the Appendices.


\section{One-dimensional model}
\label{sec:model}

In this paper we address the problem of spin pumping in
low-dimensional materials in contact with a FM where a precessing
magnetization is induced. In such systems, itinerant electrons travel
from the NM portion into the FM with a random spin orientation and
back. The magnetization of FM changes the orientation of the spin of
the returning electrons, and angular momentum leaks out of the FM and
into the NM region as a spin current. To model such a hybrid FM/NM
system, the FM region can be viewed as a time-dependent boundary
condition to the NM region.

\begin{figure}[h]
  \centering
  \includegraphics{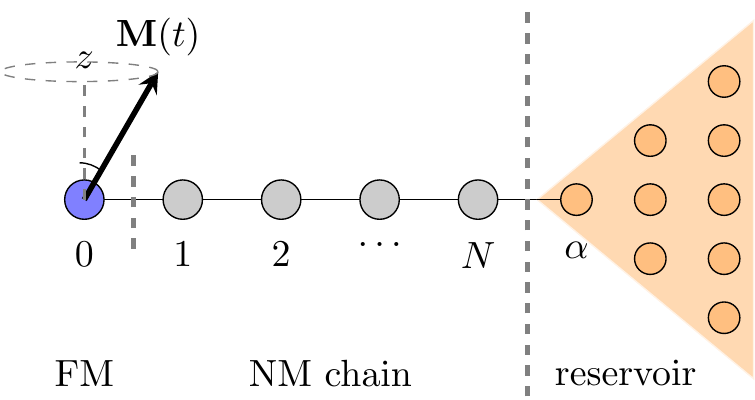}
  \caption{Scheme of the one-dimensional model of spin pumping from a
    magnetic site representing a ferromagnet (FM) to a nonmagnetic
    (NM) chain connected to a reservoir.}
  \label{fig:1Dsystem}
\end{figure}

We begin by considering the idealized situation of a one-dimensional
system, see Fig. \ref{fig:1Dsystem}. We adopt the transport
formulation developed by Dhar and Shastry \cite{shastry} as the
starting point and extend it to include spin-dependent and
time-dependent boundary conditions in the special case of a single
reservoir attached to the nonmagnetic metal region.

Consider a one-dimensional chain where the site at $j=1$ is connected
to a magnetic site at $j=0$ as shown in Fig. \ref{fig:1Dsystem}. At
the magnetic site, itinerant electrons interact with the
time-dependent magnetization of the FM,
\begin{equation}
\mb{M}(t) = M_\parallel \hat{\mb{z}} + M_\perp \left( \hat{\mb{x}}
\cos\Omega t - \hat{\mb{y}} \sin\Omega t \right).
\label{eq:magnetization}
\end{equation}
The dynamics of the magnetization is determined by the
Landau-Lifshitz-Gilbert equation, where a damping term is introduced
phenomenologically to account for magnetization loses.\cite{spc-book}
Here, we assume that Eq. (\ref{eq:magnetization}) describes the
stationary state of the magnetization and includes any damping. The
opposite end of the chain, at the site $j=N$, is connected to a
reservoir via a site $\alpha$. A hopping term describes the itinerant
electronic motion along the chain, where no spin-orbit mechanism is
present at this point. The Hamiltonian of each segment reads
\begin{equation}
\hl_{\rm mag} = -\frac{J}{2}\, \mb{M}(t) \cdot \sum_{s,s'} a_{s}^\dag\,
\bs{\sigma}_{ss'}\, a_{s'},
\label{eq:H_M}
\end{equation}
\begin{eqnarray}
\hl_{\rm chain} & = & -\sum_{j = 1}^{N-1} \sum_{s,s'} \left(
c_{j+1,s}^\dag\, \tau_{j;s,s'}\, c_{j,s'} + c_{j,s}^\dag\,
\tau_{j;s',s}^\ast \, c_{j+1,s'} \right) \nn \\ & & +\ \sum_{j=1}^N
\sum_{s} V_{j,s}\ c_{j,s}^\dagger\, c_{j,s},
\end{eqnarray}
and
\begin{equation}
\hl_{\rm res} = -\sum_{\lambda,\eta} \sum_{s} T_{\lambda\eta}\,
d_{\lambda,s}^\dag d_{\eta,s},
\end{equation}
where $s,s'=\uparrow,\downarrow$. The fermionic operators $a_{s}$,
$c_{j,s}$, and $d_{\lambda,s}$ act on the magnetic, chain, and
reservoir sites, respectively and obey the standard anticommutation
relations. $\bs{\sigma} = (\sigma^x,\sigma^y,\sigma^z)$ are Pauli
matrices. The parameters $\tau_{j;s,s'}=\tau_{j;s',s}^\ast$ describe
the hopping amplitude between neighboring sites $j$ and $j+1$ in the
chain and could be spin dependent; in the absence of spin-orbit
coupling, $\tau_{j;s,s'} = \delta_{s,s'}\, \tau_j$. The on-site
potential $V_{j,s}$ is included to account for inhomogeneities in the
chain. Finally, the matrix elements $T_{\lambda\eta}$ describe the
site connectivity in the reservoir, which can be complex.

The coupling between the magnetic site and the chain and between the
chain and the reservoir are assumed spin independent and are given by
the Hamiltonians
\begin{equation}
  \hl_{\rm mag-chain} = -\gamma_0 \left( a_{s}^\dag\, c_{1,s} +
  c_{1,s}^\dag\, a_{s} \right)
\end{equation}
and
\begin{equation}
  \hl_{\rm chain-res} = -\gamma_\alpha \left( c_{N,s}^\dag\,
  d_{\alpha,s} + d_{\alpha,s}^\dag\, c_{N,s} \right),
\end{equation}
respectively.


\subsection{Equations of Motion}
\label{sec:equations-motion}

Equations of motion for the fermionic particle operators are obtained
using the standard Heisenberg equation of motion, e.g.,
$\dot{c}_{j,s}= i[\hl,c_{j,s}]$, where
\begin{equation}
\hl = \hl_{\rm mag} + \hl_{\rm chain} + \hl_{\rm res} + \hl_{\rm
  mag-chain} + \hl_{\rm chain-res}
\end{equation}
(we assume $\hbar=1$). To simplify the notation, the time-dependent
and time-independent amplitudes in Eq. (\ref{eq:H_M}) resulting after
the insertion of Eq. (\ref{eq:magnetization}) can be cast as frequency
parameters $\Omega_\| = -\frac{J}{2}M_\|$ and $\Omega_\perp =-J
M_\perp$. We then obtain
\begin{eqnarray}
\label{eq:motion_t}
\dot{\ab}(t) & = & -i \Omega_\| \sigma_z \ab(t) - i
\Omega_\perp\, \left( \sigma^+ e^{i\Omega t} + \sigma^- e^{-i\Omega t}
\right) \ab(t) \nn \\ & & +\ i\gamma_0\, \cb_{1}(t)
\end{eqnarray}
for the magnetic site and
\begin{equation}
\dot{\cb}_{1}(t) = -i{\bf V}_1\, \cb_{1}(t) + i\gamma_0\, \ab(t) +
i\bs{\tau}_1\, \cb_{2}(t),
\end{equation}
\begin{equation}
\dot{\cb}_{j}(t) = -i{\bf V}_j\, \cb_{j}(t) + i\bs{\tau}_{j-1}\,
\cb_{j-1}(t) + i\bs{\tau}_j\, \cb_{j+1}(t),
\end{equation}
with $1<j<N$, and
\begin{equation}
\dot{\cb}_{N}(t) = -i{\bf V}_N\, \cb_{N}(t) + i\bs{\tau}_{N-1}\,
\cb_{N-1}(t) + i\gamma_\alpha\, \db_{\alpha}(t)
\end{equation}
for the chain sites. In the expressions above, we introduced the
spinor particle operators $\ab =
\begin{pmatrix}
  a_{\up} \\ a_{\down}
\end{pmatrix}$,
$\cb_j =
\begin{pmatrix}
  c_{j,\up} \\ c_{j,\down}
\end{pmatrix}$,
and
$\db_\alpha =
\begin{pmatrix}
  d_{\alpha,\up} \\ d_{\alpha,\down}
\end{pmatrix}$
and the matrices $\bs{\tau} =
\begin{pmatrix}
  \tau_{\uparrow,\uparrow} & \tau_{\uparrow,\downarrow} \\
  \tau_{\downarrow,\uparrow} & \tau_{\downarrow,\downarrow}
\end{pmatrix}$
and
${\bf V}_j =
\begin{pmatrix}
  V_{j,\uparrow} & 0 \\
  0 & V_{j,\downarrow}
\end{pmatrix}$.

For the equations of motion of the reservoir operators, we get
homogeneous equations for the bulk and an equation containing an
inhomogeneous term due to the coupling to the chain,
\begin{equation}
\dot{\db}_{\eta}(t) = i\sum_\nu T_{\eta \nu}\, \db_{\nu}(t), \qquad
\eta\neq \alpha,
\label{eq:c_gen}
\end{equation}
and
\begin{equation}
\label{eq:c_alpha}
\dot{\db}_{\alpha}(t) = i\gamma_\alpha\, \cb_{N}(t) + i\sum_\nu
T_{\alpha \nu}\, \db_{\nu }(t).
\end{equation}

Combining Eqs. (\ref{eq:c_gen}) and (\ref{eq:c_alpha}), we can express
the general solution for the operator of the site $\alpha$ with spin
state $s$ in the integral form
\begin{eqnarray}
d_{\alpha,s}(t) & = & i\sum_\eta g^r_{\alpha\eta}(t-t_0)\, d_{\eta,s}(t_0)
\nn \\ & & -\ \gamma_\alpha \int_{t_0}^\infty g_{\alpha\alpha}^r(t-t')\, c_{N,s}(t')\,
dt',
\label{eq_alpha}
\end{eqnarray}
where the homogeneous part of the solution,
\begin{equation}
\label{eq:h_s}
h_s(t) = i\sum_\eta g^r_{\alpha\eta}(t-t_0)\, d_{\eta,s }(t_0),
\end{equation}
plays the role of a noise-like term and the inhomogeneous part in
Eq. (\ref{eq_alpha}) is dissipative in nature.\cite{shastry} In
Eqs. (\ref{eq_alpha}) and (\ref{eq:h_s}), $g_{\lambda\eta}^r$ denotes
the retarded Green's function of the decoupled reservoir and reads
\begin{equation}\label{eq:gr_res}
g_{\lambda\eta}^r(t-t') = -i\theta(t-t') \sum_n \phi_n^\ast(\lambda)\,
\phi_n(\eta)\, e^{-iE_n (t-t')},
\end{equation}
where $\{\phi_n\}$ are the single-particle eigenfunctions of the
reservoir with eigenenergy $\{E_n\}$ (see Appendix
\ref{sec:reservoir}).

In the following, we assume that at a time $t=t_0$ the reservoir is in
thermal equilibrium, such that
\begin{equation}
\left\langle d^\dagger_{n,s} (t_0)\, d_{n',s'} (t_0) \right\rangle =
\delta_{n,n'}\, \delta_{s,s'}\, f(E_n),
\label{eq:thermal_eq}
\end{equation}
where $d_{n,s}(t) = \sum_\lambda d_{\lambda,s} \phi_n(\lambda)$,
$f(\varepsilon) = 1/[e^{(\varepsilon-\mu)/T}+1]$ is the Fermi-Dirac
distribution, and $T$ and $\mu$ and the reservoir's temperature and
chemical potential, respectively (we assume $k_B=1$).


\subsection{Fourier Transform of the Equations of Motion}
\label{sec:four-transf}

It is useful to express the equations of motion in frequency
domain. For that purpose, let us use the following convention for the
Fourier transform of the particle operators and other time-dependent
terms:
\begin{equation}
a_{s}(t) = \int
  \frac{d\omega}{2\pi} a_{s}(\omega)\, e^{-i\omega t},
\end{equation}
\begin{equation}
c_{j,s}(t) = \int
  \frac{d\omega}{2\pi} c_{j,s}(\omega)\, e^{-i\omega t},
\end{equation}
\begin{equation}
d_{\lambda,s}(t) = \int
  \frac{d\omega}{2\pi} d_{\lambda,s}(\omega)\, e^{-i\omega t},
\end{equation}
\begin{equation}
h_s(t) = \int \frac{d\omega}{2\pi} h_s(\omega)\, e^{-i\omega t},
\end{equation}
and
\begin{equation}
g^r_{\lambda\eta}(t) = \int \frac{d\omega}{2\pi}
g^r_{\lambda\eta} (\omega)\, e^{-i\omega t}.
\end{equation}
Inserting these definitions into Eqs. (\ref{eq:motion_t}) to
(\ref{eq:h_s}), we obtain
\begin{eqnarray}
\label{eq_motion_omega}
& & (\omega - \Omega_\| \sigma_z)\, \ab(\omega) - \int d\omega'\,
\hl_1(\omega,\omega')\, \ab(\omega') \nn \\ & & = -\ \gamma_0\,
\cb_1(\omega),
\end{eqnarray}
\begin{equation}
\omega\, \cb_1(\omega) = {\bf V}_1\, \cb_1(\omega) - \gamma_0\,
\ab(\omega) - \tau_1\, \cb_2(\omega),
\end{equation}
\begin{equation}
\omega\, \cb_j(\omega) = {\bf V}_j\, \cb_j(\omega) - \bs{\tau}_{j-1}\,
\cb_{j-1}(\omega) - \bs{\tau}_j\, \cb_{j+1}(\omega),
\end{equation}
with $1<j<N$,
\begin{equation}
\omega\, \cb_N(\omega) = {\bf V}_N\, \cb_N(\omega) - \bs{\tau}_{N-1}\,
\cb_{N-1}(\omega) - \gamma_\alpha\, \db_\alpha(\omega),
\end{equation}
and
\begin{eqnarray}
\label{eq:motion_c_alpha}
\db_\alpha(\omega) & = & \mathbf{h}(\omega) - \gamma_\alpha\,
g_{\alpha\alpha}^r(\omega)\, \cb_N(\omega),
\end{eqnarray}
where the Fourier transform of the time-dependent part of the Hamiltonian
is given by the expression
\begin{equation}
\label{eq:hNrelation}
\hl_1(\omega,\omega') = \Omega_\perp [ \sigma^+ \delta(\omega' -
  \omega + \Omega) + \sigma^- \delta(\omega' - \omega - \Omega)],
\end{equation}
with $\sigma^\pm = (\sigma^x \pm i \sigma^y)/2$, and $\mathbf{h}
= \begin{pmatrix} h_\uparrow \\ h_\downarrow \end{pmatrix}$. Notice
that $\hl_1$ is a $2\times 2$ matrix in spin space.


\section{Charge and Spin Currents}
\label{sec:charge-spin-current}

The expression for the charge current follows from the continuity
equation in a discrete one-dimensional lattice,
\begin{equation}
  \frac{\partial \rho_j}{\partial t} + \left( J_{j+1}^c - J_{j}^c \right)
  = 0,
\end{equation}
where $\rho_j = \cb_j^\dagger \cb_j$ is the charge density operator at
the site $j$ (both the electron charge and the lattice constant are
assumed to be unity). Using the equation of motion for $\cb_j$, the
particle current operator between sites $j-1$ and $j$ can be cast as
\begin{equation}
  J_j^c(t) = i \left[ \cb_j^\dag(t)\, \bs{\tau}_{j-1}\, \cb_{j-1}(t) -
    \cb_{j-1}^\dag(t)\, \bs{\tau}_{j-1}\, \cb_j(t) \right].
\label{eq:charge_current}
\end{equation}

Let us first consider the case when no spin-orbit coupling is present
in the chain, namely, when $\bs{\tau}$ is diagonal. Equation
(\ref{eq:charge_current}) gives us the total charge current as a sum
of spin up and down currents at the site $j$. However, to obtain the
local spin current we need to keep in mind that when an electron with
spin up is moving to the left, it produces an effect equivalent to an
electron with spin down moving to the right as far as the transfer of
angular momentum is concerned. In both cases, up spin angular momentum
is transferred to the right. A general expression for spin continuity
can be introduced by using the rate of change of magnetization and the
conservation of angular momentum,\cite{spc-book}
\begin{equation}
\label{eq:spin_continuity}
  \frac{\partial \bs{s}_j}{\partial t} + \left( \mb{J}_{j+1} -
  \mb{J}_{j} \right) = 0,
\end{equation}
where the spin density at the site $j$ is defined as $\bs{s}_j =
\frac{1}{2} \cb_j^\dag\, \bs{\sigma}\, \cb_j(t)$ and $\mb{J}_j$ is the
spin current operator between sites $j-1$ and $j$,\cite{note}
\begin{equation}
  \mb{J}_j(t) = \frac{i}{2} \left[ \cb_j^\dag(t)\, \bs{\sigma}\,
    \bs{\tau}_{j-1}\, \cb_{j-1}(t) - \cb_{j-1}^\dag(t)\,
    \bs{\tau}_{j-1}\, \bs{\sigma}\, \cb_j(t) \right],
\label{eq:J_j}
\end{equation}
which is Hermitian: $\left[\mb{J}_j(t)\right]^\dagger = \mb{J}_j(t)$.

Now let us consider the case when there is spin-orbit coupling in the
chain. In general, an external torque acting on the spin density at
each site has to be included. The source torque can be due to on-site
spin scattering process or to spin-orbit terms that cannot be reduced
to the divergence of a current. Equation (\ref{eq:J_j}) still holds
for a system with spin-orbit interactions, but an extra source torque
term due to on-site spin scattering processes is needed in the
continuity equation (\ref{eq:spin_continuity}), which must be replaced
by
\begin{equation}
\label{eq:spin_continuity_spin-orbit}
\frac{\partial \bs{s}_j}{\partial t} + \left( \mb{J}_{j+1} -
\mb{J}_{j} \right) = {\bf T}_{j},
\end{equation}
where the torque at site $j$ is defined as
\begin{equation}
{\bf T}_j = \frac{i}{2} \cb_j^\dagger\, \left[ \bs{\sigma}, {\bf
    V}_j \right]\, \cb_j.
\end{equation}
In Ref. \onlinecite{SpinCurrent-definition} it was pointed out that
the proper definition of spin current at the macroscopic level
requires adding a contribution from the local external torque, such
that Eq. (\ref{eq:spin_continuity}) is restored. In other words, the
external torque must be absorbed into the current expression. However,
the microscopic nature of our model enables us to distinguish between
the transfer of angular momentum either as spin currents or as a
conversion to the other degrees of freedom. Therefore, we will adopt
Eq. (\ref{eq:J_j}) even when spin-orbit coupling is present. In fact,
the proper definition of the spin current in the presence of
spin-dependent processes has been a source of debate in the literature
\cite{rashba-sd, tokatly-sd, sonin1-sd, sonin2-sd}. One aspect that
makes the definition nontrivial is the existence of intrinsic
nondissipative background currents. In such systems, even without any
dynamical source of current or spin chemical potential difference, a
spin current can flow. As Sonin \cite{sonin1-sd, sonin2-sd} pointed
out, regardless of the definition of the spin current, a source torque
term is needed to compensate for the tranfer of spin angular to
orbital angular momentum. In this paper we adopt Eq. (\ref{eq:J_j}) as
the spin current expression. We return to discuss this definition in
Sec. \ref{sec:finite-chain-so} when we deriving an expression for the
current in the presence of spin-orbit interaction.

The Fourier transform of the spin current between sites $j-1$ and $j$
of the chain takes the form
\begin{eqnarray}
\label{eq:J_general}
 \mb{J}_{j}(\omega) & = & \frac{i}{2} \int \frac{d\omega'}{2\pi}
 \left[ \cb_j^\dag(\omega')\, \bs{\sigma}\, \bs{\tau}_{j-1}\,
   \cb_{j-1}(\omega'+\omega) \right. \nn \\ & &
   \left. -\ \cb_{j-1}^\dag(\omega'-\omega)\, \bs{\tau}_{j-1}\,
   \bs{\sigma}\, \cb_j(\omega') \right].
\end{eqnarray}
Notice that, in Fourier space, the current is no longer Hermitian;
instead, it satisfies $\left[\mb{J}_j(\omega)\right]^\dagger =
\mb{J}_j(-\omega)$. In particular, the $z$ components of the current
can be written as
\begin{equation}
\label{eq:J_jz}
J_j^z(\omega) = J_{j,\uparrow}(\omega) - J_{j,\downarrow}(\omega),
\end{equation}
where
\begin{eqnarray}
  \label{eq-spins-current}
  J_{j,s}(\omega) & = & \frac{i}{2} \sum_{s'} \tau_{j-1;s,s'}\, \int
  \frac{d\omega'}{2\pi} \left[ c_{j,s}^\dag(\omega')\,
    c_{j-1,s'}(\omega'+\omega) \right. \nn \\ & & \left. -\ \eta_{s}
    \eta_{s'}\, c_{j-1,s}^\dag(\omega'-\omega)\, c_{j,s'}(\omega')
    \right],
\end{eqnarray}
and $\eta_{\uparrow,\downarrow} = \pm 1$.

Because of the harmonic nature of the precessing magnetization at the
$j=0$ site, the expectation value of the Fourier transform of the spin
current can be cast as a sum over multiples of the oscillation
frequency $\Omega$, namely,
\begin{equation}
\label{eq:J_decomp}
\leg \mb{J}_j (\omega) \rg = 2\pi \sum_k \mb{I}_j(\omega_k)\,
\delta(\omega - \omega_k),
\end{equation}
where $\omega_k = k\, \Omega$ and $k$ is an integer. The stationary
(dc) spin current can then be directly related to the zeroth harmonic
component,
\begin{eqnarray}
\overline{\leg \mb{J}_j (t) \rg} & \equiv & \lim_{T\rightarrow\infty}
\frac{1}{T} \int_{t}^{t+T} dt'\, \leg \mb{J}_j (t') \rg \\ & = &
\sum_k \mb{I}_j(\omega_k) \lim_{T\rightarrow\infty} \frac{
  e^{-i\omega_k(t+T/2)}\sin\left(\frac{\omega_kT}{2}\right)}
    {\omega_kT/2} \\ & = & \mb{I}_j(0).
\label{eq:Js_dc}
\end{eqnarray}
%


\section{Spin Pumping in the Absence of a Chain}
\label{sec:zero_chain}

For the sake of simplicity, we first evaluate the spin current for the
case $N=0$, when the reservoir is directly connected to the magnetic
site. The study of the zero-length chain gives us some insight into
the behavior of spin pumping currents and serves to guide us in
derivations involving finite-length chains. Following
Eq. (\ref{eq-spins-current}), the spin-$s$ component of current in
Fourier space reads (the site index can be dropped)
\begin{eqnarray}
\label{eq_current_zerol}
  J_s(\omega) & = & \frac{i\gamma}{4\pi} \int d\omega' \left[
    d_{\alpha,s}^\dag (\omega')\, a_{s}(\omega'+\omega) \right. \nn
    \\ & & \left. -\ a^\dag_{s}(\omega'-\omega)\,
    d_{\alpha,s}(\omega') \right],
\end{eqnarray}
where $\gamma=\gamma_0=\gamma_\alpha$. The equations of motion for the
chainless case can be obtained from Eqs. (\ref{eq_motion_omega}) and
(\ref{eq:motion_c_alpha}),
\begin{eqnarray}
& & (\omega - \Omega_\| \sigma^z)\, \ab(\omega) - \int d\omega'\,
  \hl_1(\omega,\omega')\, \ab(\omega') \nn \\ & = & -\ \gamma\,
  \db_\alpha(\omega),
\label{eq_nzero}
\end{eqnarray}
and
\begin{eqnarray}
\label{eq:c_alpha_zero}
  \db_\alpha(\omega) & = & \mathbf{h}(\omega) - \gamma\,
  g_{\alpha\alpha}^r(\omega)\, \ab(\omega).
\end{eqnarray}

We can use Eq. (\ref{eq:c_alpha_zero}) to eliminate $d_{\alpha,s}$
from the expression of the spin-$s$ component of the current,
$J_s(\omega) = J_{j,s}(\omega)$, by replacing $c_{j+1,s}$ with
$d_{\alpha,s}$ and $c_{j,s}$ with $a_{s}$ in
Eq. (\ref{eq-spins-current}),
\begin{eqnarray}
J_s(\omega) & = & \frac{i\gamma}{2} \int d\omega'\, \left[ h_s^\dag
  (\omega')\, a_{s}(\omega'+\omega) \right. \nn \\ & &
  \left. -\ a_{s}^\dag(\omega'-\omega)\, h_s(\omega') \right] \nn \\ &
& - i\gamma^2 \int d\omega'\, a_{s}^\dag(\omega')\,
a_{s}(\omega'+\omega) \nn \\ & & \times\, \left\{
g_{\alpha\alpha}^{a}(\omega') - g_{\alpha\alpha}^{r}
(\omega'+\omega)\right\},
\end{eqnarray}
recalling that $[g_{\alpha\alpha}(\omega)]^\ast =
g^a_{\alpha\alpha}(\omega)$. We can also substitute
Eq. (\ref{eq:c_alpha_zero}) into the the right-hand side of
Eq. (\ref{eq_nzero}) to get
\begin{eqnarray}
& & \int d\omega' \left\{\omega\, \sigma^0\, \delta(\omega-\omega')
  \right. \nn \\ & & \left. -\ \left[\hl_0 + \hl_1 + \Sigma^r
    \right](\omega,\omega') \right\} \ab(\omega') = -\gamma\,
  \mb{h}(\omega),
\label{eq_nonhmgns}
\end{eqnarray}
where the static and the dynamic parts of the Hamiltonian are
\begin{equation}
\hl_0(\omega,\omega') = \Omega_\| \sigma^z\, \delta(\omega-\omega')
\end{equation}
and
\begin{equation}
\hl_1(\omega,\omega') = \Omega_\perp[\sigma^+\,
  \delta(\omega-\omega'-\Omega) + \sigma^-\,
  \delta(\omega-\omega'+\Omega)],
\end{equation}
respectively. The self energy due to the reservoir is given by
\begin{equation}
\Sigma^r(\omega,\omega') = \gamma^2 g_{\alpha\alpha}^r(\omega)\,
\sigma^0\, \delta(\omega-\omega')
\label{eq:self-energy}
\end{equation}
and $\sigma^0$ denotes the identity operator in spin space. Further
simplification is possible by treating the right-hand side of
Eq. (\ref{eq_nonhmgns}) as a nonhomogeneous term and by writing the
magnetic-site particle operator in terms of the fully-dressed Green's
function of that site,
\begin{equation}
  a_{s}(\omega) = - \gamma \sum_{s'} \int d\omega'\,
  G_{ss'}^{r}(\omega,\omega')\, h_{s'}(\omega'),
  \label{eq:c0_h}
\end{equation}
where
\begin{eqnarray}
  & & \int d\omega'' \left\{\omega\, \sigma^0\, \delta(\omega-\omega'')
  - \left[\hl_0 + \hl_1 + \Sigma^r \right](\omega,\omega'') \right\}
  \nn \\ & & \times\, G^r(\omega'',\omega') = \sigma^0\,
  \delta(\omega-\omega').
\label{eq:GF}
\end{eqnarray}
Thus, we can express the magnetic-site operator $c_{0,s}$ entirely in
terms of the noise-like operator $h_s$. In the limit of
$t_0\rightarrow -\infty$, it is possible to show that the correlation
function for $h_s(\omega)$ is diagonal in spin and frequency (see
Appendix \ref{sec:init-therm-equil}),
\begin{equation}
  \leg h_s^\dag(\omega)\, h_{s'}(\omega') \rg = \delta_{s,s'}\,
  \delta(\omega' - \omega)\, I_\alpha(\omega),
  \label{eq:h_corr}
\end{equation}
where $I_\alpha(\omega) = \rho_\alpha(\omega)f(\omega)$ and
$\rho_\alpha(\omega)$ is the reservoir's density of states at the site
$\alpha$,
\begin{eqnarray}
\label{eq:DOS}
\rho_\alpha(\omega) & = & -\frac{1}{\pi}\, \Im \left[
  g_{\alpha\alpha}^r(\omega) \right] \\ & = & \sum_n
|\phi_n(\alpha)|^2\delta(\omega -E_n).
\end{eqnarray}
Using Eqs. (\ref{eq:h_corr}) and (\ref{eq:c0_h}), one arrives at the
following expression for the expectation value of the spin-$s$
component of the current:
\begin{eqnarray}
  \leg J_s(\omega) \rg & = & \frac{i\gamma^2}{2} \int d\omega'\,
  \left\{ \mathcal{F}_{s}(\omega,\omega') +
  \mathcal{I}_{s}(\omega,\omega') \right. \nn \\ & & \left.  \times
  \left\{ g_{\alpha\alpha}^r(\omega'+\omega) -
  g_{\alpha\alpha}^{a}(\omega') \right\} \right\},
\label{eq:J_s}
\end{eqnarray}
where $\mathcal{F}$ and $\mathcal{I}$ are functions of the
magnetic-site Green's functions $G^{r,a}$, with $G^a = \left(
G^r\right)^\dagger$,
\begin{eqnarray}
  \mathcal{F}_s(\omega,\omega') & = & \left[
    G_{ss}^{a}(\omega',\omega'-\omega) -
    G_{ss}^r(\omega'+\omega,\omega') \right] \nn \\ & &
  \times\ I_\alpha(\omega')
\end{eqnarray}
and
\begin{eqnarray}
\mathcal{I}_s(\omega,\omega') & = & \gamma^2 \int d\omega''\sum_{s'}
G_{ss'}^{a}(\omega'',\omega') G_{ss'}^r(\omega'+\omega,\omega'') \nn
\\ & & \times\, I_\alpha(\omega'').
\end{eqnarray}

As we argue in Sec. \ref{sec:pertubation}, from the perturbative
expansion of the Green's function in powers $\Omega_\perp$, we know
that even terms are diagonal in both spin and frequency, while odd
terms are only nonzero when they involve opposite spin
indices. Therefore, in general, one can write
\begin{equation}
G_{ss}(\omega,\omega') = \delta(\omega-\omega')\, D_{s}(\omega),
\end{equation}
leading to
\begin{equation}
\mathcal{F}_s(\omega,\omega') = \delta(\omega)\, \Im \left[
  D_{s}^r(\omega') \right] I_\alpha(\omega').
\end{equation}
It is then useful to rewrite $\mathcal{I}_s$ in terms of
same-spin-state and opposite-spin-state Green's functions, namely,
\begin{eqnarray}
\label{eq:I}
  \mathcal{I}_s(\omega,\omega') & = & \gamma^2 \int d\omega'' \left[
    G_{ss}^{a} (\omega'',\omega')\, G_{ss}^r (\omega'+\omega,\omega'')
    \right. \nn \\ & & \left. +\, G_{s\bar{s}}^{a}
    (\omega'',\omega')\, G_{s\bar{s}}^r (\omega'+\omega,\omega'')
    \right] I_\alpha(\omega'').
\end{eqnarray}
Using the Green's function relation
\begin{equation}
  G^r - G^a = G^r\, \left[ \Sigma^r - \Sigma^a \right]\, G^a,
  \label{eq:Gdifference}
\end{equation}
it is possible to show that the first term in the integrand on the
right-hand side of Eq. (\ref{eq:I}) cancels $\mathcal{F}_s$ exactly,
leading to
\begin{eqnarray}
\leg J_s(\omega) \rg & = & -\frac{i\gamma^4}{2}\, \int d\omega' \int
d\omega'' \nn \\ & & \times\, G^{a}_{\bar{s}s}(\omega'',\omega')\,
G_{s\bar{s}}^r(\omega'+\omega,\omega'') \nn \\ & & \times\,
I_\alpha(\omega'') \left[g^a_{\alpha\alpha}(\omega') -
  g^r_{\alpha\alpha}(\omega'+\omega) \right],
\label{eq_zfinal}
\end{eqnarray}
which is the central result of this Section.

Following similar steps, one can derive expressions for the other spin
components of the current. The results can be combined into a single
expression that generalizes Eq. (\ref{eq:J_s}), namely, 
\begin{eqnarray}
\leg \mb{J}(\omega) \rg & = & \frac{i\gamma^2}{2} \int
\frac{d\omega'}{2\pi} \left\{ \mb{F}(\omega,\omega') +
\mb{I}(\omega,\omega') \right. \nn \\ & & \left. \times \left[
  g_{\alpha\alpha}^r(\omega'+\omega) - g_{\alpha\alpha}^{a}(\omega')
  \right] \right\},
  \label{eq_general_exp}
\end{eqnarray}
where
\begin{eqnarray}
\mb{F}(\omega,\omega') & = & \sum_{s,s'} \bs{\sigma}_{ss'} \left[
  G_{s's}^{a}(\omega',\omega'-\omega) -
  G_{s's}^{r}(\omega'+\omega,\omega') \right] \nn \\ & & \times\,
I_\alpha(\omega')
\label{eq:Fgen}
\end{eqnarray}
and
\begin{eqnarray}
\mb{I}(\omega,\omega') & = & \gamma^2 \int d\omega'' \sum_{s,s',s_1}
G_{s_1s}^{a}(\omega'',\omega') \nn \\ & & \times\ \bs{\sigma}_{ss'}
G_{s's_1}^r(\omega'+\omega,\omega'')\, I_\alpha(\omega'').
\label{eq:Igen}
\end{eqnarray}
%


\subsection{Perturbative Expansion in $\Omega_\perp$} 
\label{sec:pertubation}

In most situations of experimental
relevance,\cite{ando-spinInjection,ando-sI2} the transverse amplitude
of time-dependent field driving the magnetization precession in the FM
is much smaller than the longitudinal static component, resulting in
$\Omega_\perp \ll \Omega_\|$. We consider this regime and expand the
magnetic-site Green's function in powers of $\Omega_\perp$, namely, in
powers of the time-dependent Hamiltonian term $\hl_1$:
\begin{equation}
\label{eq_G_expansion}
G = G^{(0)} + G^{(0)}\, \hl_1\, G^{(0)} + G^{(0)}\, \hl_1\, G^{(0) }\,
\hl_1\, G^{(0)} + \ldots.
\end{equation}

The zeroth-order (static) magnetic-site Green's function $G^{(0)}$ is
obtained by solving Eq. (\ref{eq:GF}) when $\hl_1$ is absent, yielding
\begin{equation}
G^{(0)}_{ss'}(\omega,\omega') = \delta_{s,s'}\,
\delta(\omega-\omega')\, {\cal G}_s(\omega),
\end{equation}
where
\begin{equation}
{\cal G}_{s}(\omega) = \frac{1}{\omega - \eta_s \Omega_\| - \gamma^2
  g_{\alpha\alpha}(\omega)}
\end{equation}
and $\eta_{\uparrow,\downarrow} = \pm 1$. Thus, the zeroth-order
Green's function is diagonal in spin space.

The first-order Green's function has only off-diagonal spin terms,
\begin{eqnarray}
G_{\up\up}^{(1)}(\omega,\omega') & = & 0,
\\ G_{\up\down}^{(1)}(\omega,\omega') & = & \Omega_\perp
\delta(\omega'-\omega-\Omega)\, \gup(\omega)\, \nn \\ & &
\times\ \gdn(\omega+\Omega), \\ G_{\down\up}^{(1)}(\omega,\omega') & =
& \Omega_\perp \delta(\omega'-\omega+\Omega)\, \gdn(\omega)\, \nn \\ &
& \times\ \gup(\omega-\Omega), \\ G_{\down\down}^{(1)}(\omega,\omega')
& = & 0,
\end{eqnarray}
while the second-order Green's function recovers the spin-diagonal
structure of the zeroth-order case,
\begin{eqnarray}
G_{\up\up}^{(2)}(\omega,\omega') & = &
\Omega_\perp^2\delta(\omega'-\omega)\, \gup(\omega)\, \gup(\omega)
\\ & & \times\, \gdn(\omega+\Omega),
\\ G_{\up\down}^{(2)}(\omega,\omega') & = & 0,
\\ G_{\down\up}^{(2)}(\omega,\omega') & = & 0,
\\ G_{\down\down}^{(2)}(\omega,\omega') & = & \Omega_\perp^2
\delta(\omega'-\omega)\, \gdn(\omega)\, \gdn(\omega) \nn \\ & &
\times\, \gup(\omega-\Omega).
\end{eqnarray}

The spin dependence of higher order contributions to the Green's
function repeats this pattern: diagonal for even orders and
off-diagonal for odd orders. In addition, even orders are also
diagonal in the frequency variables.


\subsection{Spin Current Components}
\label{sec:z-component}

From the final expression for the spin-$s$ state component of the
current, Eq. (\ref{eq_zfinal}), and the expansion of the Green's
function up to second order in $\Omega_\perp$, one finds the following
expression for the $z$-component of the spin current:
\begin{eqnarray}
\leg J^z(\omega) \rg & = & \delta(\omega)\, \pi \gamma^4\,
\Omega_\perp^2 \int d\omega' \rho_\alpha(\omega') \nn \\ & & \times
\left[ \left| {\cal G}_\up^{r}(\omega') \right|^2 \left| {\cal
    G}_\down^{r}(\omega'+\Omega) \right|^2 I_\alpha(\omega'+\Omega)
  \right. \nn \\ & & \left. -\ \left| {\cal G}_\down^{r}(\omega')
  \right|^2 \left| {\cal G}_\up^{r}(\omega'-\Omega) \right|^2
  I_\alpha(\omega'-\Omega) \right] \nonumber \\ & &
+\ O(\Omega_\perp^4).
\label{eq_zcomp_final}
\end{eqnarray}
Since only the zero-frequency component is nonzero, upon returning to
the time representation and utilizing Eq. (\ref{eq:Js_dc}), this
relation yields a nonzero dc current, namely,
\begin{eqnarray}
\overline{\leg J^z(t) \rg} & = & \frac{\gamma^4\, \Omega_\perp^2}{2}
\int d\omega\, \rho_\alpha \left( \omega - \Omega/2 \right)\,
\rho_\alpha \left( \omega + \Omega/2 \right) \nonumber \\ & & \times
\left| {\cal G}_\up^{r} \left(\omega - \Omega/2 \right) \right|^2
\left| {\cal G}_\down^{r} \left( \omega + \Omega/2 \right) \right|^2
\nonumber \\ & & \times \left[ f \left( \omega + \Omega/2 \right) - f
  \left( \omega - \Omega/2 \right) \right] \nonumber \\ & &
+\ O(\Omega_\perp^4),
\label{eq:zcomp_dc}
\end{eqnarray}
where we have symmetrized the frequency integrand for convenience.

We notice that inverting the static magnetic field and the direction
of precession (e.g., $\Omega\rightarrow -\Omega$ and
$\Omega_\|\rightarrow - \Omega_\|$) flips the spin of the zeroth-order
Green's function ${\cal G}_\up(\omega) \rightarrow {\cal
  G}_\down(\omega)$. As a result, the spin current reverses its
direction. This is expected on the basis of time-reversal
symmetry. Moreover, at zero precession or zero transverse magnetic
field, the spin current vanishes.



Considering now the $x$ component of the integral $\mb{F}$ in
Eq. (\ref{eq:Fgen}), we obtain
\begin{eqnarray}
F^{x} & = & -\gamma \left[ G^a_{\down\up}(\omega',\omega'-\omega) -
  G^r_{\down\up}(\omega'+\omega,\omega') \right. \nn \\ & &
  \left. +\ G^a_{\up\down}(\omega',\omega'-\omega) -
  G^r_{\up\down}(\omega'+\omega,\omega') \right. \nn \\ & & \left.
  \times\ I_\alpha(\omega') \right].
\end{eqnarray}
Notice that all terms contain opposite-spin-state Green's functions,
thus vanish in even powers in $\Omega_\perp$ but are
$\Omega$-dependent in odd powers of $\Omega_\perp$. As a result, in
the time domain, $F^x$ oscillates and, upon averaging over one
precession period, it vanishes. A similar argument can be used to show
that $I^x$ vanishes as well. Therefore, all transverse components of
the spin current vanish when averaged over time.


\subsection{Interface Parameters}
\label{sec:interface-parameter}

The dynamics of the FM magnetization in the adiabatic approximation is
governed by the Landau-Lifshitz-Gilbert (LLG) equation,
\begin{equation}
\label{eq:LLG}
  \frac{d\mb{m}}{dt} = \gamma\, \mb{m}\times \mb{H}_{\rm eff} +
  \alpha\, \mb{m} \times \frac{d\mb{m}}{dt},
\end{equation}
where $\mb{m}$ is the magnetization unit vector, $\gamma$ is the
gyromagnetic ratio, $\mb{H}_{\rm eff}$ is the effective magnetic field
(including the external magnetic field and the local demagnetization
field), and $\alpha$ is the Gilbert damping parameter. In the absence
of any contact between the FM and a NM, the relaxation of the
magnetization occurs entirely through processes internal to the FM,
which are phenomenologically accounted for by the parameter
$\alpha$. When a NM is brought in contact with the FM, the
magnetization relaxation can also happen through angular momentum
leaking into the NM as a spin current. To account for this
contribution, consider that the effective magnetic field applied to
the FM to be of the form
\begin{equation}
  \mb{H}_{\rm eff} = h_x(t)\, \hat{\mb{x}} + h_y(t)\,
  \hat{\mb{y}} + H_\|\, \hat{\mb{z}},
\end{equation}
where $H_\|$ is the static component of the field while $h_x$ and
$h_y$ are the time-dependent components. Following the scattering
theory of spin pumping,\cite{Brataas_Spin_battery} the spin current
can be expressed as
\begin{equation}
\label{eq:gmix-scurrent}
\mb{I}_{\rm spin} = \frac{1}{4\pi}\, g_{\up\down}\,
\mb{m}\times\frac{d\mb{m}}{dt},
\end{equation}
where the mixing conductance $g_{\up\down}$ is defined in terms of
reflection matrices as
\begin{equation}
  g^{\up\down} = \sum_{m,n} \left( \delta_{m,n} - r^\up_{mn}
  r^\down_{mn} \right),
\end{equation}
with the sum taken over transverse conducting channels. Notice the
similarity of the right-hand side of Eq. (\ref{eq:gmix-scurrent}) with
the the Gilbert damping term in Eq. (\ref{eq:LLG}). One can absorb the
angular momentum leakage contribution on the magnetization relaxation
due to the spin current by substituting $\alpha$ with $\alpha^\prime$
in Eq. (\ref{eq:LLG}), where
\begin{equation}
\label{eq:alphaprime}
\alpha^\prime = \alpha + \frac{g_L\, A_r}{4\pi M}.
\end{equation}
Here, $g_L$ is the Land\'{e} factor, $M$ is the total (bulk)
magnetization of the FM, and $A_r = \Re {g^{\up\down}}$ (in most
practical situations, the imaginary component of the mixing
conductance can be neglected).

In the small precessing field approximation, $h_\perp =
\sqrt{h_x^2+h_y^2} \ll |H_\parallel|$, one can solve the LLG equation
for the stationary solution of the dynamics of magnetization to get
\begin{equation}
m_\perp(t) = |m_\perp|\, e^{-i(\Omega t + \delta)},
\end{equation}
where
\begin{equation}
|m_\perp| = \frac{\gamma M h_\perp}{\sqrt{ (\alpha' M \Omega)^2 +
    (\gamma H_\| + \Omega)^2 } }
\end{equation}
and
\begin{equation}
\tan\delta = \frac{\alpha' M \Omega}{\gamma H_\| + \Omega}.
\end{equation}
After substituting $m_\perp(t)$ in Eq. (\ref{eq:gmix-scurrent}), we
arrive at
\begin{equation}
\label{eq:Ispin_scatt}
I_{\rm spin}^z = \frac{1}{4\pi} \Omega\, {|m_\perp|}^2 g_{\up\down}.
\end{equation}
We can combine this expression with that obtained in
Sec. \ref{sec:z-component} for the spin current in terms of the
system's Green's function, Eq. (\ref{eq:zcomp_dc}) to obtain an
expression for the mixing conductance in terms of Green's functions,
\begin{equation}
g^{\up\down} = \frac{\pi J^2 \gamma^4}{2\hbar} \int d\omega
\rho_\alpha^2(\omega) \left| {\cal G}_\up^{r}(\omega) \right|^2 \left|
    {\cal G}_\down^{r}(\omega) \right|^2 \frac{df(\omega)}{d\omega}.
\label{eq:g-updown}
\end{equation}

In experiments, there are two standard approaches to
quantify the spin pummping current and both are
indirect. The first and most common consists of measuring
the broadening of the FMR spectrum and utilizing
Eqs. (\ref{eq:alphaprime}) and
(\ref{eq:Ispin_scatt}).\cite{mosendz_pump,
  ando-sI2,heinrich-fmr1} The second is to infer the current
magnitude through the observation of the {\em inverse spin
  Hall effect} (ISHE) in the NM when a sufficiently strong
spin-orbit coupling is present.\cite{saitoh_pump,
  mosendz-ishe, ando-ishe} Although, the latter seems more
direct, the relation between the measured ISHE voltage and
the actual spin current depends on various materials
parameters which are often not accurately
known.\cite{ishe-measurement} Equation (\ref{eq:g-updown})
provides a useful relation between the physical properties
of medium where the spin current that is generated propagates to
the enhanced broadening of FMR due to the angular momentum
leakage. When generalized to higher dimensions,
Eq. (\ref{eq:g-updown}) provides a recipe for {\em ab initio}
calculations of the Gilbert parameter.


\section{Spin Pumping with a Finite Chain}
\label{sec:finite-chain}

The formulation developed for the $N=0$ chain in
Sec. \ref{sec:zero_chain} can be extended to a finite-length
chain. The equivalent to the equation of motion (\ref{eq_nonhmgns})
for the particle operators in the chain can be written as
\begin{equation}
\label{eq:eq_motion_Z}
  \sum_{j'=0}^N \sum_{s'} \int d\omega'
  \mz^r_{j,s;j',s'}(\omega,\omega')\, c_{j',s'}(\omega') =
  -\gamma_\alpha\, \delta_{j,N} h_{s}(\omega),
\end{equation}
where $0\leq j\leq N$ and we introduced $c_{0,s}\equiv a_{s}$. The
matrix $\mz^r$ can be split into two contributions,
\begin{equation}
\mz^r = \mz_0^r + \mz_1^r,
\end{equation}
where
\begin{widetext}
\begin{eqnarray}
\left[\mz_0^r\right]_{j,s;j',s'}(\omega,\omega') & = & \delta_{s,s'}\,
\delta(\omega-\omega') \left\{ \delta_{j,j'}\, \delta_{j,0}\, \left[
  (\omega - \Omega_\|)\, \delta_{s,\up} + (\omega+\Omega_\|)\,
  \delta_{s,\down} \right] + (\omega - V_{j,s})\, \delta_{j,j'} -
\delta_{j,j'}\, \delta_{j,N}\, \gamma_\alpha^2\,
g_{\alpha\alpha}^r(\omega) \right\} \nn \\ & &
+\ \delta(\omega-\omega')\, \left( \delta_{j,j'+1}\, \tau_{j-1;s,s'} +
\delta_{j,j'-1} \, \tau_{j;s,s'} \right).
\end{eqnarray}
and
\begin{equation}
\left[\mz_1^r\right]_{j;j'}(\omega,\omega') = \delta_{j,0}\,
\delta_{j',0}\, \Omega_\perp \left[ \sigma^+
  \delta(\omega'-\omega-\Omega) + \sigma^- \delta(\omega' - \omega +
  \Omega) \right].
\end{equation}
\end{widetext}

Let us define the retarded Green's function of the finite chain as
$G^r \equiv \left(\mz^r \right)^{-1}$. We can then solve
Eq. (\ref{eq:eq_motion_Z}) for the particle operator and write
\begin{equation}
\label{eq_pr_op}
  c_{j,s}(\omega) = -\gamma_\alpha \sum_{s'} \int d\omega'\,
  G^r_{j,s;N,s'} (\omega,\omega')\, h_{s'}(\omega'),
\end{equation}
where $0\leq j\leq N$. The Green's function can be expanded in powers
of $\Omega_\perp$ similarly to Eq. (\ref{eq_G_expansion}). Since
$\mz^{(0)}$ is diagonal in frequency, one can write the zeroth order
term as
\begin{equation}
G_{j,s;j';s'}^{(0)}(\omega,\omega') = \delta(\omega - \omega') \,
{\cal G}_{\sbt{j,s;j',s'}}(\omega).
\end{equation}
Using this expression, the first-order contribution is found to be
\begin{widetext}
\begin{equation}
G_{j,s;j',s'}^{(1)}(\omega,\omega') = \Omega_\perp \left[ {\cal
    G}_{j,s;0,\uparrow}(\omega)\, {\cal
    G}_{0,\downarrow;j',s'}(\omega+\Omega)\,
  \delta(\omega'-\omega-\Omega) + {\cal
    G}_{j,s;0,\downarrow}(\omega)\, {\cal
    G}_{0,\uparrow;j',s'}(\omega-\Omega)\,
  \delta(\omega'-\omega+\Omega) \right].
\end{equation}
Similarly, for the second-order contribution we have
\begin{eqnarray}
G_{j,s;j',s'}^{(2)}(\omega,\omega') & = & \Omega_\perp^2 \left[
  \delta(\omega'-\omega-2\Omega)\, {\cal G}_{j,s;0,\up}(\omega)\,
        {\cal G}_{0,\down;0,\up}(\omega+\Omega)\, {\cal
          G}_{0,\down;j's'}(\omega') \right. \nonumber \\ & &
        \left. +\ \delta(\omega'-\omega+2\Omega)\, {\cal
          G}_{j,s;0,\down}(\omega)\, {\cal
          G}_{0,\up;\down,0}(\omega-\Omega)\, {\cal
          G}_{0,\up;j's'}(\omega') \right. \nonumber \\ & &
        \left. +\ \delta(\omega'-\omega)\, {\cal
          G}_{j,s;0,\up}(\omega)\, {\cal
          G}_{0,\down;0,\down}(\omega+\Omega)\, {\cal
          G}_{0,\up;j's'}(\omega') \right. \nonumber \\ & &
        \left. +\ \delta(\omega'-\omega)\, {\cal
          G}_{j,s;0,\down}(\omega)\, {\cal
          G}_{0,\up;0,\up}(\omega-\Omega)\, {\cal
          G}_{0,\down;j's'}(\omega') \right].
\end{eqnarray}
Notice that in the absence of spin-orbit coupling in the chain, ${\cal
  G}_{0,\down;\up,0} = {\cal G}_{0,\up;\down,0} = 0$ and the inelastic
(off diagonal in frequency) contribution to the second-order Green's
function vanishes.
\end{widetext}


\subsection{Current in the presence of spin-orbit coupling}
\label{sec:finite-chain-so}

If electrons experience no spin scattering in the chain, the spin
$s$-state current flows homogeneously from the magnetic site, along
the chain, and into the reservoir without spin-orbit coupling. Thus,
it can be shown that the spin current will remain the same as
Eq. (\ref{eq:zcomp_dc}).

When spin-orbit is present, the spin current will vary along the
chain. In this case, one is required to use Eq. (\ref{eq:J_general})
to compute the three components of the spin current at a given site
$j$. Let us focus on the $z$ component. Substituting
Eq. (\ref{eq_pr_op}) and its Hermitian conjugate into
Eq. (\ref{eq:J_general}), we obtain
\begin{widetext}
\begin{eqnarray}
J_j^z(\omega) & = & \frac{i\gamma_\alpha^2}{4\pi} \int d\omega' \int
d\omega'' \int d\omega''' \nonumber \\ & & \times\,
\mathbf{h}^\dagger(\omega'') \left[ \bs{G}^a_{N;j}
  (\omega'',\omega')\, \sigma^z \bs{\tau}_{j-1}\,
  \bs{G}^r_{j-1;N}(\omega'+\omega,\omega''') - \bs{G}^a_{N;j-1}
  (\omega'',\omega'-\omega)\, \bs{\tau}_{j-1}\, \sigma^z\,
  \bs{G}^r_{j;N}(\omega',\omega''') \right] \mathbf{h}(\omega'''), \nn
\\ & &
\label{eq:J_z_spin}
\end{eqnarray}
where $0\leq j \leq N$ and $\bs{G}^{r(a)}_{j;j'}$ denotes the $2\times
2$ retarded (advanced) Green's function connecting sites $j$ and
$j'$. Using the correlation function introduced in
Eq. (\ref{eq:h_corr}), we can take the expectation value of
Eq. (\ref{eq:J_z_spin}) to obtain
\begin{eqnarray}
\left\langle J_j^z(\omega) \right\rangle & = &
\frac{i\gamma_\alpha^2}{4\pi} \int d\omega' \int d\omega'' \nonumber
\\ & & \times\, {\rm tr} \left[ \bs{G}^a_{N;j} (\omega'',\omega')\,
  \sigma^z \bs{\tau}_{j-1}\, \bs{G}^r_{j-1;N}(\omega'+\omega,\omega'')
  - \bs{G}^a_{N;j-1} (\omega'',\omega'-\omega)\, \bs{\tau}_{j-1}\,
  \sigma^z\, \bs{G}^r_{j;N}(\omega',\omega'') \right]
I_\alpha(\omega''). \nn \\ & &
\label{eq:J_z_spin_expec}
\end{eqnarray}
\end{widetext}
where the trace is over spin variables. Equation
(\ref{eq:J_z_spin_expec}) is one of the main results of this paper. It
provides a framework for computing the $z$ component of the spin
current at any site within the chain that connects the magnetic site
and the reservoir. Unfortunately, any further simplification of this
expression is daunting. Similarly to the case where the reservoir is
connected directly to the magnetic site, Sec. \ref{sec:pertubation},
we can use the perturbative expansion of the Green's function in
powers of $\Omega_\perp$. The result is still rather involved if the
spin-dependent hopping amplitude $\bs{\tau}$ is kept general and is
not presented here.

A more compact expression can be obtained for the spin current between
the last site of the chain and the reservoir, even in the presence of
a general spin-orbit hopping amplitude. For that purpose, we take a
step back, set $j=\alpha$ in Eq. (\ref{eq:J_general}), and consider
the $z$ component of the spin current operator,
\begin{eqnarray}
\label{eq:J_alpha_z}
J_{\alpha}^z(\omega) & = & \frac{i\gamma_\alpha}{4\pi} \int d\omega'
\sum_s \eta_s \Big[ d_s^\dag(\omega')\, c_{N,s}(\omega'+\omega) \nn
  \\ & & \-\ c_{N,s}^\dag(\omega'-\omega)\, d_s(\omega') \Big].
\end{eqnarray}
Using Eqs. (\ref{eq:motion_c_alpha}) and
(\ref{eq_pr_op}), taking the expectation value, and using
Eq. (\ref{eq:h_corr}), we can rewrite Eq. (\ref{eq:J_alpha_z}) as
\begin{widetext}
\begin{eqnarray}
\label{eq:Jalpha}
\left \langle J_{\alpha}^z(\omega) \right\rangle & = & -
\frac{i\gamma_\alpha^2}{4\pi} \int d\omega' \sum_s \eta_s \Big\{
I_\alpha(\omega') \left[ G_{N,s;N,s}^r(\omega'+\omega,\omega') -
  G_{N,s;N,s}^a(\omega',\omega'-\omega) \right] \nn \\ & &
-\ \gamma_\alpha^2 \int d\omega'' \sum_{s'} I_\alpha (\omega'') \left[
  g^r_{\alpha\alpha}(\omega') - g_{\alpha\alpha}^a(\omega'+\omega)
  \right] G_{N,s';N,s}^a(\omega'',\omega')\,
G_{N,s;N,s'}^r(\omega'+\omega,\omega'') \Big\}.
\end{eqnarray}
The absence of a spin-dependent hopping amplitude in Eq.
(\ref{eq:Jalpha}) makes it more amenable to an analytical
treatment. Focusing on the dc component of the spin current, as shown
in Eqs.  (\ref{eq:J_decomp}) and (\ref{eq:Js_dc}), we expand the
Green's function harmonics of the precessing frequency $\Omega$,
namely,
\begin{equation}
G(\omega,\omega') = \delta(\omega'-\omega)\, D_0(\omega) + \sum_{k\neq
  0} \delta(\omega'-\omega - k\Omega)\, D_k(\omega).
\end{equation}
Inserting this expansion into Eq. (\ref{eq:Jalpha}) and keeping only
the terms corresponding to the dc limit, we obtain
\begin{eqnarray}
\label{eq:Jz_expans_Omega}
\left \langle J_{\alpha}^z(\omega) \right\rangle_{\rm dc} & = & -
\frac{i\gamma_\alpha^2}{4\pi} \delta(\omega) \int d\omega' \sum_s
\eta_s \Big\{ I_\alpha(\omega') \left[ D_{0;N,s;N,s}^r (\omega') -
  D_{0;N,s;N,s}^a (\omega') \right] \nn \\ & & -\ \gamma_\alpha^2
\left[ g_{\alpha\alpha}^r(\omega') - g_{\alpha\alpha}^a(\omega')
  \right] I_\alpha(\omega') \sum_{s'} D_{0;N,s;N,s'}^r (\omega')\,
D_{0;N,s';N,s}^a (\omega') \nn \\ & & -\ \gamma_\alpha^2 \left[
  g_{\alpha\alpha}^r(\omega') - g_{\alpha\alpha}^a(\omega') \right]
\sum_{k\neq 0} I_\alpha(\omega'+k\Omega)\, \sum_{s'}
D^r_{k;N,s;N,s'}(\omega')\, D^a_{-k;N,s';N,s}(\omega'+k\Omega) \Big\}.
\end{eqnarray}
We can now use the relations
\begin{equation}
G^r - G^a =\left[ {\cal Z}^r \right]^{-1} - \left[ {\cal Z}^a
  \right]^{-1} = G^r \left( Z^a - Z^r \right) G^a,
\end{equation}
where
\begin{equation}
\left[Z^a - Z^r \right]_{j,s;j',s'}(\omega,\omega') = -
\gamma_\alpha^2\, \delta_{j,j}\, \delta_{j,N}\, \delta_{s,s'}\,
\delta(\omega-\omega') \left[ g_{\alpha\alpha}^a(\omega) -
  g_{\alpha\alpha}^r(\omega) \right],
\end{equation}
to find
\begin{eqnarray}
\label{eq:D0diff}
D_{0;N,s;N,s}^r (\omega) - D_{0;N,s;N,s}^a (\omega) & = &
\gamma_\alpha^2 \left[ g_{\alpha\alpha}^r(\omega) -
  g_{\alpha\alpha}^a(\omega) \right] \sum_{s'}
D^r_{0;N,s;N,s'}(\omega)\, D^a_{0;N,s',N,s}(\omega) \nn \\ & &
+\ \gamma_\alpha^2 \sum_{k\neq 0} \left[
  g_{\alpha\alpha}^r(\omega+k\Omega) -
  g_{\alpha\alpha}^a(\omega+k\Omega) \right] \sum_{s'}
D^r_{k;N,s;N,s'}(\omega)\, D^a_{-k;N,s',N,s}(\omega+k\Omega).
\end{eqnarray}
Combing Eqs. (\ref{eq:Jz_expans_Omega}) and (\ref{eq:D0diff}),
recalling that $g_{\alpha\alpha}^r(\omega) - g_{\alpha\alpha}^a =
-2\pi i \rho_\alpha(\omega)$ and using Eq. (\ref{eq:DOS}), we arrive
at
\begin{equation}
\left \langle J_{\alpha}^z(\omega) \right\rangle_{\rm dc} =
-\frac{\gamma_\alpha^4}{2} \delta(\omega) \int d\omega' \sum_{k\neq 0}
\rho_\alpha(\omega')\, \rho_{\alpha}(\omega'+k\Omega) \left[
  f(\omega') - f(\omega'+k\Omega) \right] \sum_{s,s'} \eta_s\,
D^r_{k;N,s;N,s'}(\omega')\, D^a_{-k;N,s',N,s}(\omega'+k\Omega).
\end{equation}
Symmetrizing the frequency integration, we finally obtain the
following expression for the dc spin current at the interface with the
reservoir:
\begin{eqnarray}
\overline{{\left\langle J_\alpha^z(t) \right\rangle}} & = &
\frac{\gamma_\alpha^4}{2} \int d\omega \sum_{k>0}
\rho_\alpha(\omega+k\Omega/2)\, \rho_\alpha(\omega-k\Omega/2) \left[
  f(\omega+k\Omega/2) - f(\omega-k\Omega/2) \right] \nn \\ & &
\times\, {\rm tr} \left\{ \sigma^z\, \left[
  D^r_{k;N;N}(\omega-k\Omega/2)\, D^a_{-k;N;N} (\omega+k\Omega/2) -
  D^r_{-k;N;N}(\omega+k\Omega/2)\, D^a_{k;N;N} (\omega-k\Omega/2)
  \right] \right\},
\label{eq-sc-higherORD}
\end{eqnarray}
\end{widetext}
where the trace is over spin indices. Notice that in the limit of zero
pumping frequency ($\Omega\rightarrow 0$), the spin current goes to
zero.

At this point, we can go back to the perturbative expansion of the
Green's functions in powers of $\Omega_\perp$ and notice the
following:
\begin{eqnarray}
D_{-1;j,s;j',s'}(\omega) & = & \Omega_\perp \, {\cal
  G}_{j,s;0,\downarrow}(\omega)\, {\cal
  G}_{0,\uparrow;j',s'}(\omega-\Omega) \nn \\ & &
+\ O(\Omega_\perp^3),
\end{eqnarray}
and
\begin{eqnarray}
D_{1;j,s;j',s'}(\omega) & = & \Omega_\perp \, {\cal
  G}_{j,s;0,\uparrow}(\omega)\, {\cal
  G}_{0,\downarrow;j',s'}(\omega+\Omega) \nn \\ & &
+\ O(\Omega_\perp^3).
\end{eqnarray}
Since $D_{k} \sim O(\Omega_\perp^k)$, by keeping only the leading term
in powers of $\Omega_\perp$ we obtain
\begin{widetext}
\begin{eqnarray}
\label{eq:final_result}
\overline{{\left\langle J_\alpha^z(t) \right\rangle}} & = &
\frac{\gamma_\alpha^4\, \Omega_\perp^2}{2} \int d\omega\,
\rho_\alpha(\omega+\Omega/2)\, \rho_\alpha(\omega-\Omega/2) \left[
  f(\omega+\Omega/2) - f(\omega - \Omega/2) \right] \nn \\ & &
\times\, \sum_{s,s'} \eta_s \left[ \left| {\cal
    G}^r_{N,s;0,\uparrow}(\omega -\Omega/2) \right|^2 \left| {\cal
    G}^r_{0,\downarrow;N,s'}(\omega + \Omega/2) \right|^2 - \left|
  {\cal G}^r_{N,s;0,\downarrow}(\omega +\Omega/2) \right|^2 \left|
  {\cal G}^r_{0,\uparrow;N,s'}(\omega - \Omega/2) \right|^2 \right]
\nn \\ & & +\ O(\Omega_\perp^4).
\end{eqnarray}
It is straightforward to verify that setting $N=0$ in
Eq. (\ref{eq:final_result}) leads to Eq. (\ref{eq:zcomp_dc}). Notice
that for $\Omega \ll T,\mu$, the current is proportional to $\Omega$,
\begin{equation}
  \label{eq:sc-low-temp}
\overline{{\left\langle J_\alpha^z(t) \right\rangle}} \approx
\frac{\gamma_\alpha^4\, \Omega_\perp^2\, \Omega}{2} \int d\omega\,
\left[ \rho_\alpha(\omega) \right]^2 \left[ \frac{df(\omega)}{d\omega}
  \right] \sum_{s,s'} \eta_s \left[ \left| {\cal
    G}^r_{N,s;0,\uparrow}(\omega) \right|^2 \left| {\cal
    G}^r_{0,\downarrow;N,s'}(\omega) \right|^2 - \left| {\cal
    G}^r_{N,s;0,\downarrow}(\omega) \right|^2 \left| {\cal
    G}^r_{0,\uparrow;N,s'}(\omega) \right|^2 \right].
\end{equation}
\end{widetext}

Equations (\ref{eq-sc-higherORD}) and (\ref{eq:sc-low-temp}) are the
main results of this section. Equation (\ref{eq-sc-higherORD}) can be
employed to study dynamical spin pumping beyond the linear response
approximation. Combining Eq.~(\ref{eq:sc-low-temp}) with
Eq.~(\ref{eq:gmix-scurrent}) enables an atomistic calculation of the
macroscopic Gilbert parameter, which can be measured in FMR
experiments.

\begin{figure}[ht]
\includegraphics[width=0.5\textwidth]{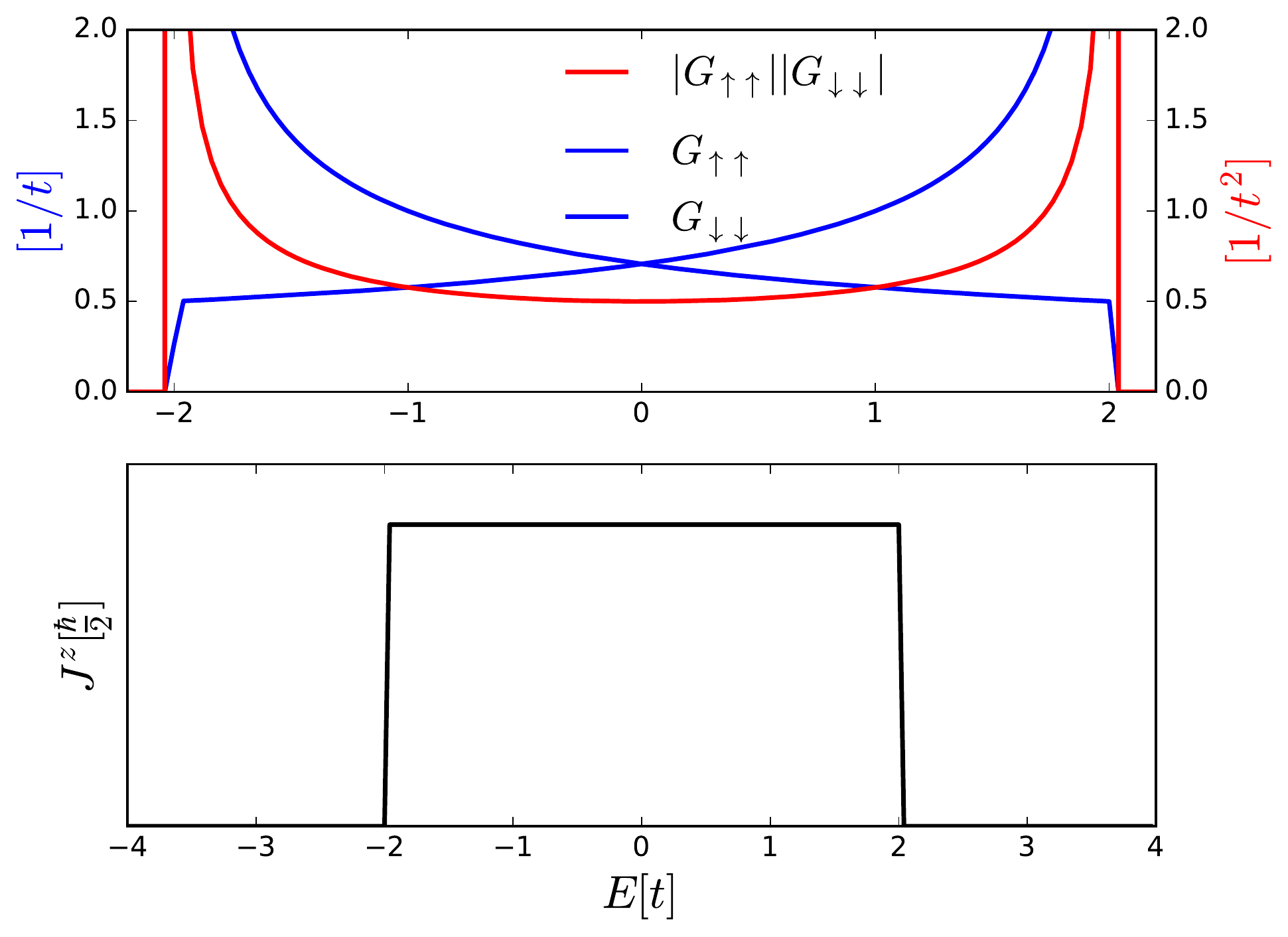}
\caption{(upper panel) Spin-diagonal components of the Green's
  function across the chain, $G_{N,s;0,s}$, in a clean (translation
  invariant) chain in the absence of spin-orbit coupling as a function
  of energy. (lower pannel) The dependence of the $z$-component of the
  spin current on the reservoir's Fermi energy. Both plots were
  obtained using parameters value such that $\gamma_\alpha^2
  \Omega_\perp^2 \Omega = 2$.}
\label{fig:clean-chain-gupup}
\end{figure}

To illustrate the results obtained so far, we performed numerical
calculations of the chain Green's function for chains of various
lengths in the presence and absence of spin-dependent on-site
potentials. In Fig. \ref{fig:clean-chain-gupup}, the spin-diagonal
components of the Green's function across the chain,
$G^{(0)}_{N,s;0,s}(E)$, and the total spin pumping current, $\langle
J^z_\alpha(E)\rangle$, are plotted as functions of energy. A constant
spin current over energy confirms that, in the absence of
spin-scattering centers, the chain is a spin-degenerate ballistic
propagating channel so long as the energy $E$ is within the energy
band. In this case, the spin current is independent of the length of
the chain.

\begin{figure}[ht]
\includegraphics[width=0.5\textwidth]{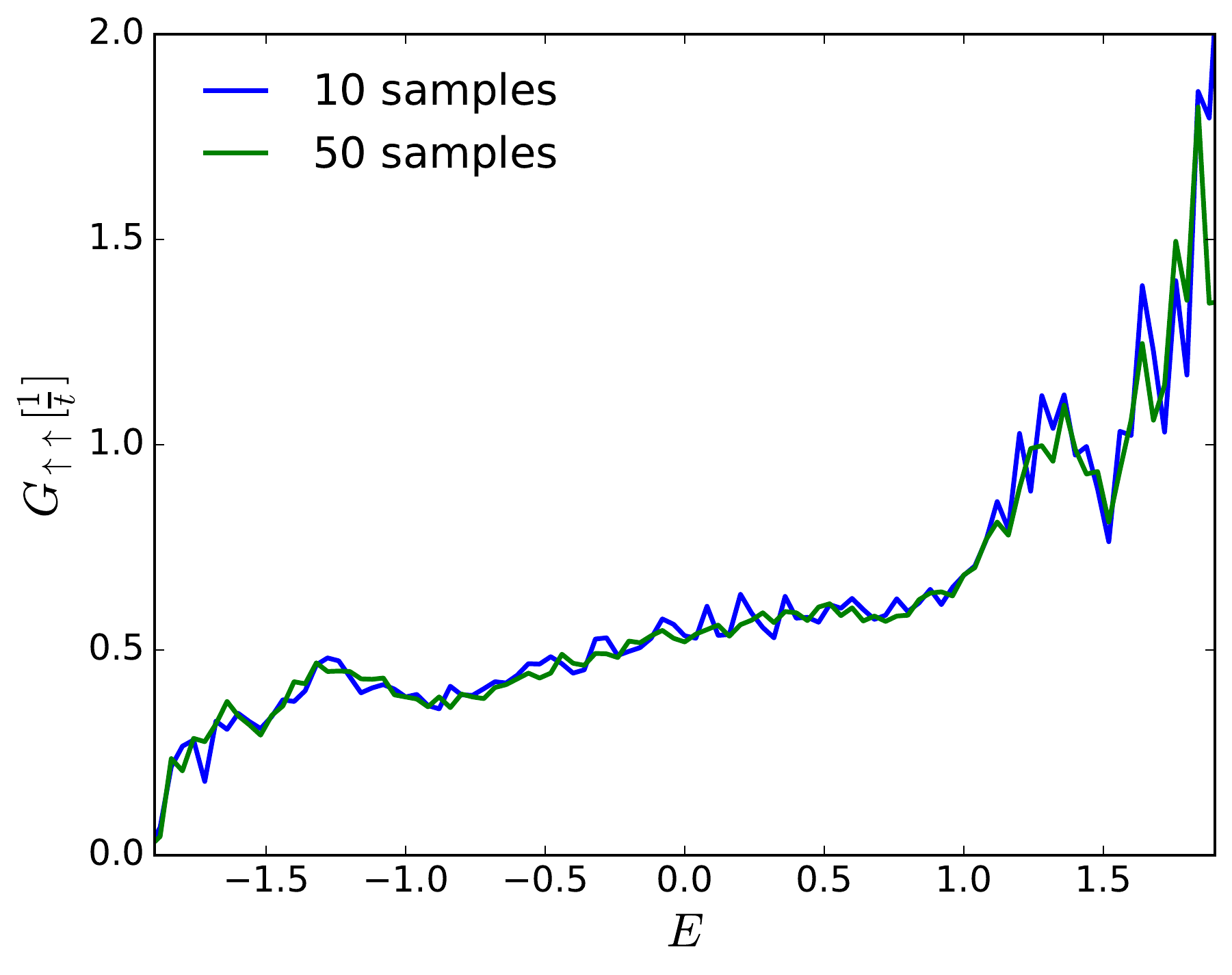}
\caption{Diagonal spin component of the Green's function across the
  chain, $G_{N,\up;0,\up}$, as a function of energy, in the presence of
  a random spin-dependent site potential. The chain length is 300
  sites and the Green's function is averaged over 10 and 50
  realizations of the random potential.}
\label{fig:samps-guu}
\end{figure}

\begin{figure}[ht]
\includegraphics[width=0.5\textwidth]{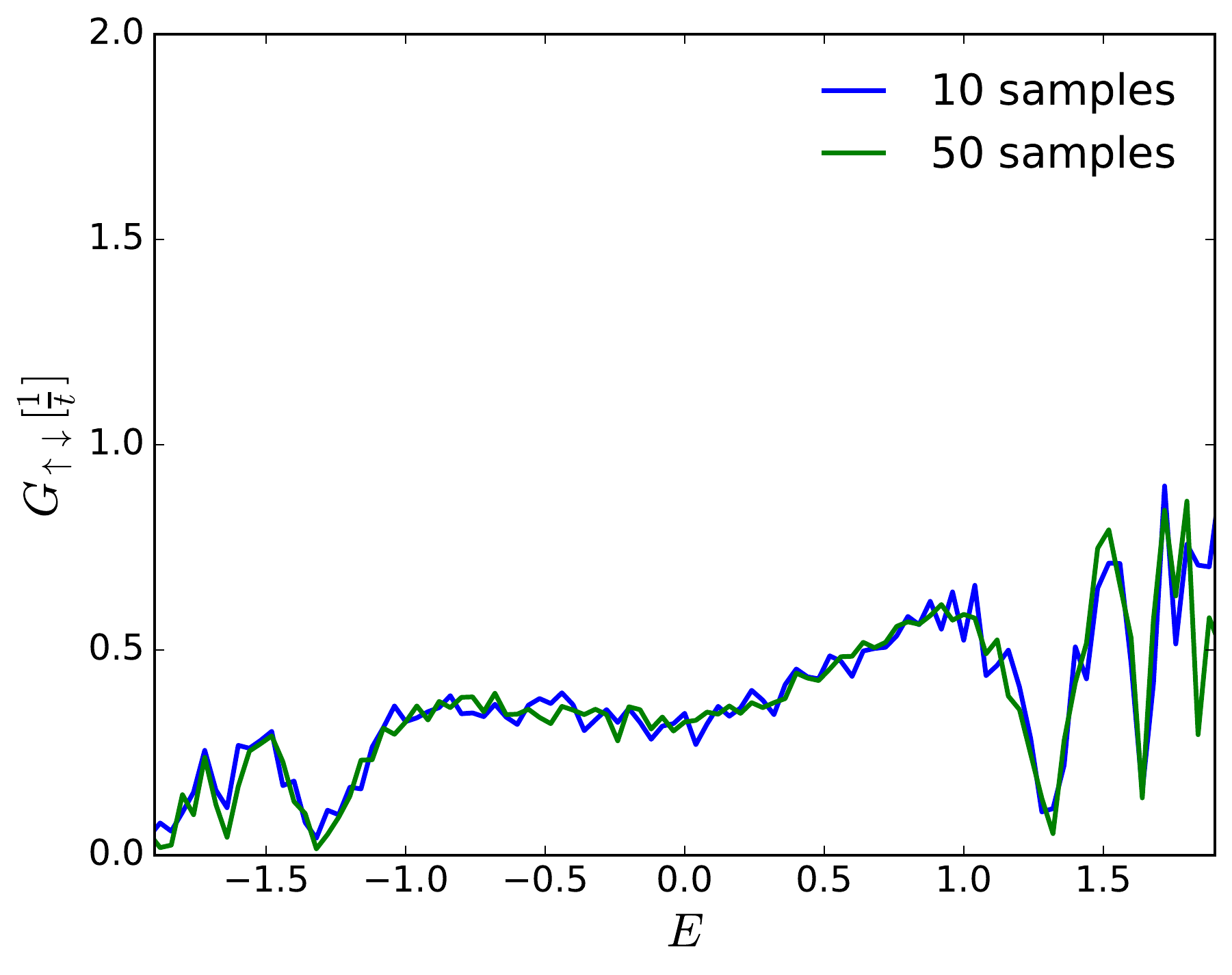}
\caption{Similarly to Fig. \ref{fig:samps-guu}, but for an off-diagonal
  spin component of the Green's function, $G_{N,\up;0,\downarrow}$.}
\label{fig:samps-gud}
\end{figure}

Figures \ref{fig:samps-gud} and \ref{fig:samps-guu} show the energy
dependence of the spin components of the chain's average Green's
function when spin-polarized impurities are introduced but no
spin-dependent hopping is present. In these simulation, $N=200$ and
$V_j = a_j^x \sigma^x + a_j^z \sigma^z$ , where the amplitudes $a_j^x$
and $a_j^z$ are randomly and uniformly chosen in the intervals
$[0,0.01 t]$ and $[0,0.05 t]$, respectively. Here $t$ denotes the
hopping amplitude in the lattice.

One of the key advantages of our formalism is that it can be utilized
to compute the relaxation of the spin current over distance from the
FM/NM interface due to spin-scattering processes in the NM region. For
large enough systems, the diffusion length can be calculated.

\begin{figure}[ht]
\includegraphics[width=0.5\textwidth]{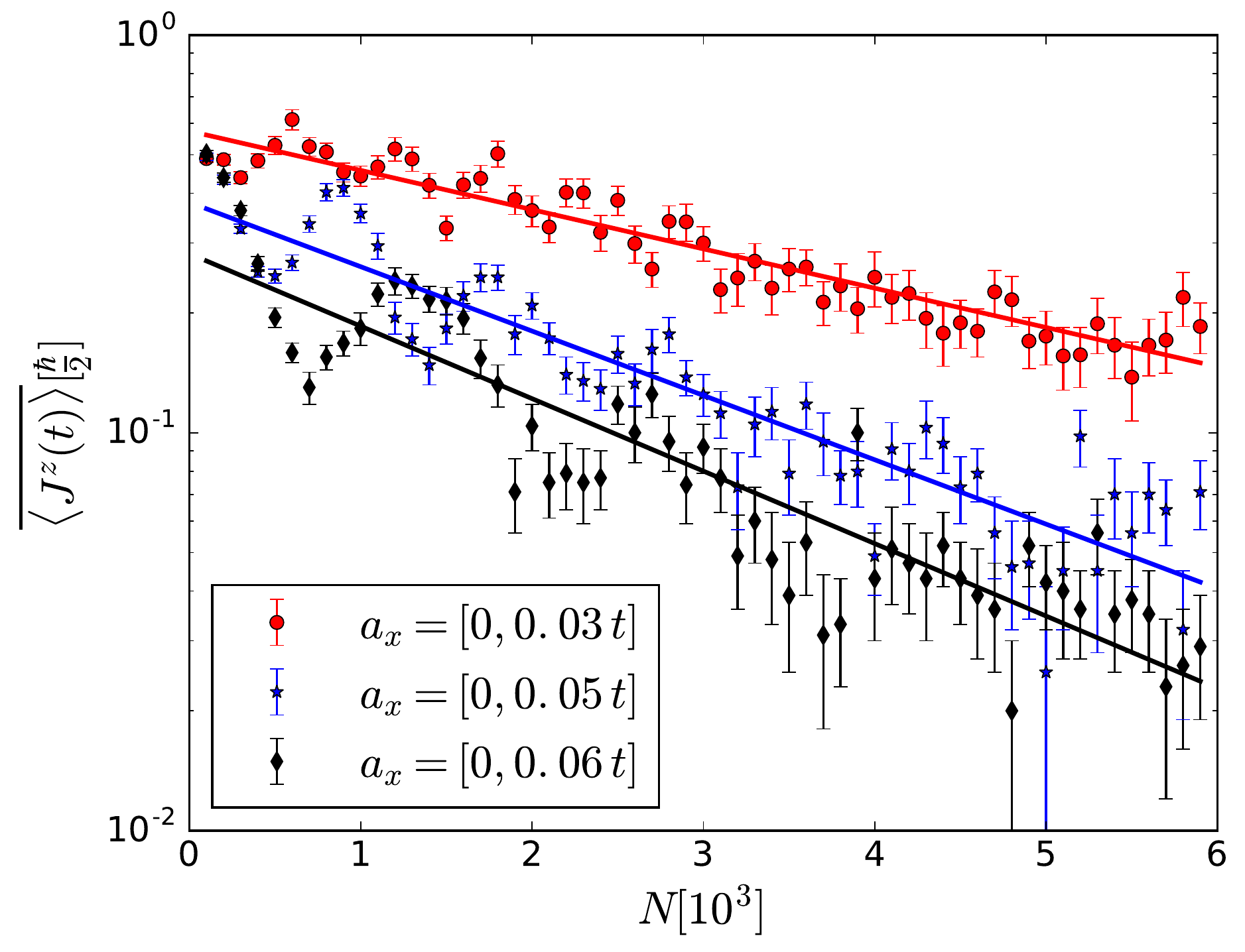}
\caption{Average dc spin current injected into the reservoir
  as a function of chain length in the presence of spin-dependent,
  random on-site potential. The data points are obtained after
  averaging over 300 samples to minimize quantum interference
  fluctuations. The solid lines are fittings to the data. The field
  $a_z$ varied within the range $[0,0.1t]$ while the $a_x$ field
  range changed for each data set, as indicated in the legend. The
  simulations are performed at $E=0$ (middle of the band).}
\label{fig:len-js}
\end{figure}

The dependence of the average dc spin pumping current on the length of
the chain is shown in Fig. \ref{fig:len-js} for the same random
spin-dependent potential. Even after averaging over 300 samples,
oscillations over the length due to interference remains. However, a
clear exponential decay emerges, with a decay length of $4.5$, $2.7$,
and $2.4$ lattice units for the three increasing disorder ranges of
$a_x$ shown in the plot.

%


\section{Extension to Two-Dimensional Systems}
\label{sec:2D-formalism}

\begin{figure}[h]
  \centering
  \includegraphics{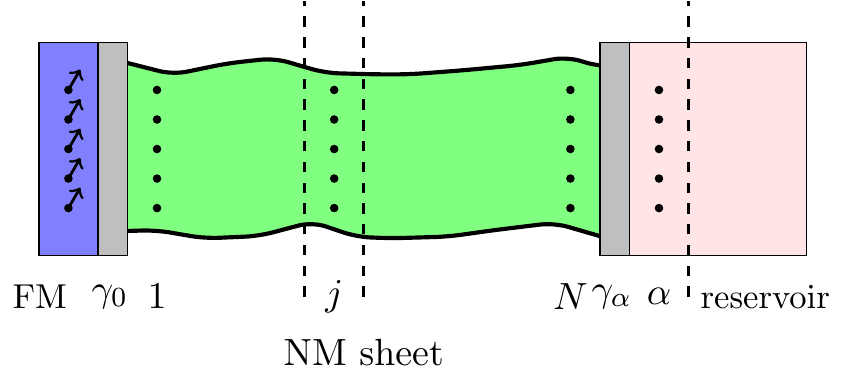}
  \caption{The two-dimensional FM/NM/reservoir system. The magnetic
    region comprises a column of sites whose magnetizations are
    synchronized. The nonmagnetic sheet is decomposed in $N$
    slices.}
\label{fig:2Dsystem}
\end{figure}

The spin pumping formulation developed in Secs. \ref{sec:model},
\ref{sec:z-component}, and \ref{sec:finite-chain} can be extended to
2D systems. To do so, we imagine the magnetic region as a column of
magnetic sites whose magnetizations precess in a synchronized way,
corresponding to a single magnetic domain. The two-dimensional
nonmagnetic region is sliced into $N$ columns and connected to a
reservoir, see Fig. \ref{fig:2Dsystem}. We keep the same notation used
for the one-dimensional finite-chain case and write the Hamiltonians
of the different regions as
\begin{equation}
  \hl_{\rm mag} = -\frac{J}{2}\, \mb{M}(t)\, \bs{a}^\dag\, \left(
  \bs{\sigma} \otimes \bs{I}_M \right)\, \bs{a}
\end{equation}
for the magnetic region,
\begin{eqnarray}
\hl_{\rm sheet} & = & -\sum_{j=1}^{N-1} \left( \bs{c}_{j+1}^\dag\,
\bs{\tau}_j\, \bs{c}_{j} + \bs{c}_{j}^\dag\, \bs{\tau}_j^\dag\,
\bs{c}_{j+1} \right) \nn \\ & & +\ \sum_{j=1}^{N} \bs{c}_{j}^\dagger\, {\bf
  V}_j\, \bs{c}_{j}
\end{eqnarray}
for the nonmagnetic region, and
\begin{equation}
\hl_{\rm res} = -\sum_{\lambda,\eta} \sum_{s} T_{\lambda\eta}\,
d_{\lambda,s}^\dag\, d_{\eta,s}
\end{equation}
for the reservoir. The Hamiltonians describing the coupling between
magnetic and nonmagnetic regions (hereafter referred to as sheet), and
between the nonmagnetic region and the reservoir are given by
\begin{equation}
\hl_{\rm mag-sheet} = -\left( \bs{a}^\dag\, \bs{\gamma}_0\, \bs{c}_{1}
+ \bs{c}_{1}^\dag\, \bs{\gamma}_0^\dag\, \bs{a} \right)
\end{equation}
and
\begin{equation}
\hl_{\rm sheet-res} = - \left( \bs{c}_{N}^\dag\, \bs{\gamma}_\alpha\,
\bs{d}_{\alpha } + \bs{d}_{\alpha}^\dag\, \bs{\gamma}_\alpha^\dag\,
\bs{c}_{N} \right),
\end{equation}
respectively, where $\bs{a}^\dag = \begin{pmatrix} a_{1} &
  a_{2} & \ldots & a_{M} \end{pmatrix}$ is the particle
operator at the column containing the magnetic region
($j=0$), $\bs{\gamma}_0$ is a $2L\times 2L$ matrix that
describes the coupling between the magnetic region and the
sheet, $\bs{c}_j^\dag = \begin{pmatrix} c_{j,1} & c_{j,2} &
  \ldots & c_{j,d_j} \end{pmatrix}$ is the particle operator
at the $j$th sheet slice, which is connected to the
neighboring $j+1$-th slice by the matrix $\bs{\tau}_i$,
$d_j$ is the number of sites in $j$th slice, and
$\bs{\gamma}_\alpha$ is the coupling matrix between the $N$th
sheet slice and the reservoir. Finally, the particle
operator acting on the sites in the reservoir that are connected
directly to the sheet is given by $\bs{d}_\alpha^\dag
= \begin{pmatrix} d_{\alpha,1} & d_{\alpha,2} & \ldots &
  d_{\alpha,d_\alpha} \end{pmatrix}$.

The equations of motion read
\begin{eqnarray}
\dot{\bs{a}}(t) & = & i\Omega_\| (\sigma_z\otimes I_M)\, \bs{a}(t)
\nn \\ & & +\ i\Omega_\perp \left[ (\sigma^+\otimes I_M)\, e^{i\Omega t} +
  (\sigma^-\otimes I_M)\, e^{-i\Omega t} \right] \bs{a}(t) \nn \\ & & +\
i\bs{\gamma}_0\, \bs{c}_1(t),
\end{eqnarray}
\begin{eqnarray}
\dot{\bs{c}_1}(t) & = & - i{\bf V}_1 + i\bs{\gamma}_0^\dag\, \bs{a} +
i\bs{\tau}_1\, \bs{c}_2(t), \\ & & \vdots \nn \\ \dot{\bs{c}_j}(t) & =
& - i{\bf V}_j + i\bs{\tau}_{j-1}^\dag\, \bs{c}_{j-1} + i\bs{\tau}_j\,
\bs{c}_{j+1}(t), \\ & & \vdots \nn \\ \dot{\bs{c}}_N(t) & = & - i{\bf
  V}_N + i\bs{\tau}_N^\dag\, \bs{c}_{N-1}(t) + i\bs{\gamma}_\alpha\,
\bs{d}_\alpha(t),
\end{eqnarray}
and
\begin{equation}
\dot{\bs{d}}_{\alpha}(t) = i \bs{\gamma}_\alpha\, \bs{c}_N(t) + i
\sum_\nu T_{\alpha\nu}\, \bs{d}_\nu(t).
\end{equation}
The Fourier transforms of the equations of motion result in
expressions similar those obtained in Sec. \ref{sec:model}, namely,
\begin{eqnarray}
& & \left[ (\omega\sigma_0 - \Omega_\| \sigma_z)\otimes I_M\right]
  \bs{a}(\omega) - \int \hl_1(\omega,\omega')\, \bs{a}(\omega) \nn
  \\ & = & - \bs{\tau}_M\, \bs{c}_1(\omega),
\end{eqnarray}
\begin{eqnarray}
\omega\, \bs{c}_1(\omega) & = & {\bf V}_1 -\bs{\gamma}_0^\dag\, \bs{a}
- \tau_1\, \bs{c}_2(\omega), \\ & & \vdots \nn \\ \omega\,
\bs{c}_j(\omega) & = & {\bf V}_j - \bs{\tau}_{j-1}^\dag\, \bs{c}_{j-1}
- \bs{\tau}_j\, \bs{c}_{j+1}, \\ & & \vdots \nn \\ \omega\,
\bs{c}_N(\omega) & = & {\bf V}_N -\bs{\tau}_{N-1}^\dag\, \bs{c}_{N-1}
- \bs{\gamma}_\alpha\, \bs{d}_{\alpha},
\end{eqnarray}
and
\begin{equation}
  \bs{d}_\alpha(\omega) = \bs{h}(\omega) - \bs{g}_{\alpha\alpha}^r
  \bs{\gamma}_\alpha^\dag(\omega)\, \bs{c}_N(\omega),
\end{equation}
where $\bs{h}$ is a vector with dimension of the surface sites
$\alpha$ in the reservoir and the Green's function of the decoupled
reservoir for slice $\alpha$ reads
\begin{eqnarray}
  \left[ \bs{g}^r_{\alpha\alpha} \right]_{i,i'}(t-t') & = &
  -i\theta(t-t')\sum_n \phi_n^*(\alpha_i)\, \phi_n(\alpha_{i'}) \nn
  \\ & & \times\ e^{-iE_n(t-t')}.
\end{eqnarray}

In order to expand the Green's function in powers of $\Omega_\perp$,
we notice that, in spin space,
\begin{equation}
\hl'_{j,j'} = \delta_{j,j'}\, \delta_{j,0}\, \Omega_\perp
  \begin{pmatrix}
    0 & \delta(\omega' - \omega - \Omega) \\
    \delta(\omega'-\omega + \Omega) & 0
  \end{pmatrix},
\end{equation}
which leads us to analogous relations to those derived in
Sec. \ref{sec:finite-chain} for the finite chain.

In order to calculate the spin current along the sheet, we can use an
expression identical to that introduced in
Sec. \ref{sec:charge-spin-current}, namely,
\begin{eqnarray}
J_{j}^z(\omega) & = & \frac{i}{2} \int \frac{d\omega'}{2\pi} \left[
  \bs{c}_{j}^\dag(\omega') \left( \sigma^z \otimes I_{d_{j}} \right)
  \bs{\tau}_{j-1}^\dag \,\bs{c}_{j-1}(\omega'+\omega) \right. \nn \\ &
  & \left. -\ \bs{c}_{j-1}^\dag(\omega'-\omega)\,
  \bs{\tau}_{j-1}^\dagger \left( \sigma^z \otimes I_{d_{j}} \right)
  \bs{c}_{j}(\omega') \right].
\end{eqnarray}
The only difference between this relation and Eq. (\ref{eq:J_general})
is that here there is an implicit sum over transverse sites. Using the
orthogonality relation of ${\bf h}(\omega)$ one can derive an
expression for the expectation value of the total spin current between
the $(j-1)$th and $j$th slices as
\begin{equation}
\label{eq:2D_scurrent}
\left\langle J_{j}^z(\omega) \right\rangle = \frac{i}{4\pi} \int
d\omega' \int d\omega''\, {\rm Tr}
\left[\mathcal{M}_{ss'}(\omega,\omega',\omega'') \right] \,
\end{equation}
where the trace is over spin and transvese site variables,
\begin{widetext}
\begin{align}
  \mathcal{M}_{ss'}(\omega,\omega',\omega'') &= \left[
    \bs{\gamma}_\alpha^\dag \bs{G}_{N;j}^a(\omega'',\omega)
    (\sigma^z\otimes\bs{I}_{d_j})
    \bs{\tau}_{j-1}^\dag \bs{G}_{j-1;N}^r(\omega+\omega',\omega'')
    \bs{\gamma}_\alpha \bs{I}(\omega'')\right. \nn\\
  &- \left. \bs{\gamma}_\alpha^\dag \bs{G}_{N;j-1}^a(\omega'',\omega)
    \bs{\tau}_{j-1}(\sigma^z\otimes 
    I_{d_j})  \bs{G}_{j-1;N}^r(\omega+\omega',\omega'')
    \bs{\gamma}_\alpha \bs{I}(\omega'') \right]
  \label{eq:2D-jsc-trace}
\end{align}
\end{widetext}
and $\bs{G}^{r(a)}_{j,j'}$ denotes the $2d_j \times 2d_{j'}$ retarded
(advanced) Green's function connecting the $j$ and $j'$ slices. A
detailed derivation of Eq. (\ref{eq:2D_scurrent}) is provided in
Appendix~\ref{sec:curr-expr-2d}.

Similar to the 1D chain, we can go further to calculate the current
at the chain-reservoir interface and expand the Green's function
harmonics of the precessing frequency $\Omega$,%
\begin{equation}
\bs{G}(\omega,\omega') = \delta(\omega'-\omega)\, \bs{D}_0(\omega) +
\sum_{k\neq 0} \delta(\omega'-\omega-k\Omega)\, \bs{D}_k(\omega),
\end{equation}
to derive
\begin{widetext}
  \begin{align}
    \overline{\leg J_{\alpha}^z(t) \rg} = \frac{1}{2} \int d\omega
    \sum_{k\neq 0} \left[ f(\omega+k\Omega) - f(\omega) \right] {\rm
      Tr} \left[\bs{\rho}_\alpha(\omega+k\Omega) \bs{\gamma}_\alpha
      \bs{D}_{k;N;N}^r(\omega') \bs{\gamma}_\alpha^\dag
      (\sigma^z\otimes \bs{I}_L)\, \bs{\rho}_\alpha(\omega)\,
      \bs{\gamma}_\alpha \bs{D}_{-k;N;N}^a(\omega'+ k\Omega)
      \bs{\gamma}_\alpha^\dag \right]
    \label{eq:2D-sc-taverage}
  \end{align}
\end{widetext}

Equation (\ref{eq:2D-sc-taverage}) is the central result of this
paper. This equation can be used to study dynamical spin pumping in
FM/NM setups beyond the linear response. In the slow-precession limit,
combined with Eq.~(\ref{eq:gmix-scurrent}), it gives us a recipe for
the calculation of the mixing conductance and the Gilbert parameter by
solely knowing the microscopic structure of the NM system.

The formalism developed in this section has several advantages over
the scattering formulation: (i) The detailed geometry of the FM/NM
systems and physical properties of the NM can be taken into account by
computing the appropriate Green's function. (ii) Since the final
expression for the spin current is written in terms of the surface
Green's functions, the recursive Green's function technique
\cite{eduardoRGF} can be utilized for an efficient computational
approach to the problem. (iii) Furthermore, since a spatial
representation of the system is used in this formalism, systems with
higher dimension and arbitrary geometry can be readily simulated.


\section{Summary and Discussion}
\label{sec:summary-discussion}

In this paper, we developed an atomistic model of spin pumping in
hybrid ferromagnetic heterostructures. The spin current expression is
given in terms of the Green's function of the nonmagnetic
portion. Motivated by the fact that, in experimental settings, the
time-dependent component of the driving magnetic field is small and
slow, we use a perturbative expansion to obtain a relation between the
mixing conductance and the physical properties of spin-carrying
medium. Among the advantages of this formalism are: (i) it provides a
framework for including the atomic structure and geometry of the
heterostructure, as well as local disorder and spin-orbit coupling
mechanism, (ii) it yields an expression for the spin current in terms
of Green's function, which can be computed using efficient recursive
computational methods, (iii) it allows us to model spin relaxation and
the ferromagnet-nonmagnetic metal interface, and (iv) when applied to
graphene, it is not limited to high doping.

In a future work we plan to apply this new computational tool to study
dynamical spin injection in realistic ferromagnet-graphene
heterostructures, and to extend the calculations to include a
determination of the spin-Hall voltage across the graphene channel
when spin-orbit coupling is included.


\acknowledgments

We are grateful to C. Lewenkopf and A. Ferreira for insightful
discussions. E.R.M. acknowledges the hospitality of the Instituto de
F\a'{i}sica at UFF, Brazil, where this work was initiated. This work
was supported in part by the NSF Grant ECCS 1402990.


\appendix


\section{Reservoir Green's function}
\label{sec:reservoir}

The retarded Green's function of the decoupled reservoir is
defined as
\begin{equation}
g_{\lambda\eta}^r(t,t') = -i\theta(t-t') \leg \{d_\lambda^\dag(t),
d_\eta(t')\} \rg.
\end{equation}
Expanding the field operators in terms of single-particle energy
eigenfunctions
\begin{equation}
d_\lambda(t) = \sum_n \phi_n(\lambda)\, d_n(t) = \sum_n
\phi_n(\lambda)\, e^{-iE_n t}\, d_n(0),
\end{equation}
the retarded Green's function of the reservoir can be written as
\begin{equation}
g_{\lambda\eta}^r(t,t') = -i\theta(t-t') \sum_n \phi_m^\ast(\lambda)\,
\phi_n(\eta)\, e^{iE_n (t-t')}.
\end{equation}
%


\section{Noise-like correlator}
\label{sec:init-therm-equil}

We can rewrite the correlation function of the noise-like term in
frequency space in terms of the fermionic operators in time using Eq
(\ref{eq:h_s}),
\begin{widetext}
\begin{eqnarray}
\left\langle h_s^\dagger(\omega)\, h_{s'}(\omega') \right\rangle & = &
\int_{-\infty}^{\infty} dt \int_{-\infty}^\infty dt'\, e^{-i(\omega t
  -\omega' t')}\,\left\langle h_\alpha^\dagger(t)\, h_{\alpha'}(t')
\right\rangle \\ & = & \sum_{\eta,\eta'} \int_{-\infty}^{\infty}
dt \int_{-\infty}^\infty dt'\, e^{-i(\omega t -\omega' t')}\,\left[
  g_{\alpha\eta}^r (t-t_0) \right]^\ast g_{\alpha\eta'}^r(t'-t_0)\,
\left\langle d^\dagger_{\eta,s} (t_0)\, d_{\eta',s'} (t_0)
\right\rangle.
\end{eqnarray}
After substituting the expansion of decoupled reservoir's Green's
function in terms of the reservoir's eigenfunction, Eq.
(\ref{eq:gr_res}), we get
\begin{eqnarray}
\left\langle h_s^\dagger(\omega)\, h_{s'}(\omega') \right\rangle & = &
\sum_{\eta,\lambda} \sum_{n,m} \phi_n(\alpha)\, \phi_n^*(\eta)\,
\phi_m^*(\alpha)\, \phi_m(\eta')\, \left\langle d^\dagger_{\eta,s}
(t_0)\, d_{\eta',s'} (t_0) \right\rangle \int_{t_0}^{\infty} dt\,
e^{-i(\omega -E_n)t} \int_{t_0}^\infty dt'\, e^{-i(E_m -\omega') t'}.
\end{eqnarray}
Using the reservoir's eigenfunction basis,
\begin{equation}
d_{\eta,s}(t_0) = \sum_n \phi_n(\eta)\, d_{n,s}(t_0),
\end{equation}
and the orthogonality of the reservoir's eigenfunctions, we obtain
\begin{eqnarray}
\left\langle h_s^\dagger(\omega)\, h_{s'}(\omega') \right\rangle & = &
\sum_{n,m} \phi_n(\alpha)\, \phi_m^*(\alpha) \left\langle
d^\dagger_{n,s} (t_0)\, d_{m,s'} (t_0) \right\rangle
\int_{t_0}^{\infty} dt\, e^{-i(\omega -E_n)t} \int_{t_0}^\infty dt'\,
e^{-i(E_m -\omega') t'}.
\end{eqnarray}
Using Eq. (\ref{eq:thermal_eq}) and taking the limit $t_0\rightarrow
-\infty$ we arrive at Eq. (\ref{eq:h_corr}).

For 2D systems, the correlation function $\leg
\bs{h}_{s_1}^\dag(\omega_1)\, \bs{h}_{s_2}(\omega_2) \rg$ can be
obtained in the same way:
\begin{eqnarray}
\leg h_{s_1,i_1}^\dagger(\omega_1)\, h_{s_2,i_2}(\omega_2) \rg & = &
\int_{-\infty}^{\infty} dt_1 \int_{-\infty}^{\infty} dt_2\,
e^{-i(\omega_1 t_1- \omega_2 t_2)}\, \leg h_{s_1,i_1}^\dag(t_1)\,
h_{s_2,i_2}(t_2) \rg \\ & = & \sum_{\eta_1,\eta_2}
\int_{-\infty}^{\infty} dt_1 \int_{-\infty}^{\infty} dt_2\,
e^{-i(\omega_1 t_1 - \omega_2 t_2)} \left[ g_{\alpha_{i_1}
    \eta_1}^r(t_1-t_0) \right]^* g_{\alpha_{i_2} \eta_2}^r(t_2-t_0)
\nn \\ & & \times \leg d^\dag_{\eta_1,s_1}(t_0)\, d_{\eta_2,s_2}(t_0) \rg \\ &
= & \sum_{\eta_1,\eta_2} \sum_{n_1,n_2} \phi_n (\alpha_{i_1})\,
\phi_{n_1}^*(\eta_1)\, \phi_{n_2}^*(\alpha_{i_2})\,
\phi_{n_2}^*(\eta_2)\, \leg d_{\eta_1,s_1}^\dag(t_0)\,
d_{\eta_2,s_2}(t_0) \nn \rg \\ & & \times \int_{t_0}^\infty dt_1\,
e^{-i(\omega_1 - E_{n_1})t_1} \int_{t_0}^\infty dt_2\,
e^{-i(E_{n_2}-\omega_2)t_2}.
\end{eqnarray}
Using the orthonormal set of eigenfunctions of the reservoir,
\begin{equation}
d_{\eta,s}(t_0) = \sum_n \phi_n(\eta)\, d_{n,s}(t_0),
\end{equation}
we can write
\begin{eqnarray}
\leg h_{s_1,i_1}^\dag(\omega_1)\, h_{s_2,i_2}(\omega_2) \rg & = &
\sum_{n_1,n_2} \phi_{n_1}(\alpha_{i_1})\, \phi_{n_2}^*(\alpha_{i_2})
\leg d^\dag_{n_1,s_1}(t_0)\, d_{n_2,s_2}(t_0) \rg \int_{t_0}^\infty
dt_1\, e^{-i(\omega_1 - E_{n_1})t_1} \int_{t_0}^\infty dt_2\,
e^{-i(E_{n_2}-\omega_2)t_2} \\ & = & \delta(\omega_1-\omega_2)\,
\delta_{s_1,s_2} \sum_{n_1} \phi_{n_1}(\alpha_{i_1})\,
\phi_{n_1}^*(\alpha_{i_2})\, \delta(\omega_1 - E_{n_1})
\end{eqnarray}
\end{widetext}
when we set $t_0\rightarrow\-\infty$. We finally arrive at
\begin{equation}
\leg \bs{h}_{s_1}^\dag(\omega_1) \bs{h}_{s_2}(\omega_2) \rg =
\delta_{s_1,s_2}\, \delta(\omega_1 - \omega_2)\,
\bs{I}_\alpha(\omega_1),
\end{equation}
where $\bs{I}_\alpha(\omega) = \bs{\rho}_\alpha(\omega) f(\omega)$ and
$\bs{\rho}_\alpha(\omega)$ is the density of states matrix at the
$\alpha$ slice,
\begin{equation}
\left[\bs{\rho}_\alpha\right]_{i_1,i_2} = \sum_{n_1} \sum_{n_1}
\phi_{n_1}(\alpha_{i_1})\, \phi_{n_1}^*(\alpha_{i_2})\,
\delta(\omega_1 - E_{n_1}).
\end{equation}
%


\section{$s$-component of the spin current}

Substituting Eq. (\ref{eq:c_alpha_zero}) into
Eq. (\ref{eq_current_zerol}), we obtain
\begin{widetext}
\begin{eqnarray}
\leg J_s(\omega) \rg & = & \frac{i\gamma}{2} \int
\frac{d\omega'}{2\pi} \left\langle \left\{ h_s^\dag(\omega') - \gamma
\left[ g_{\alpha\alpha}^{r}(\omega') \right]^* a_{s}^\dag(\omega')
\right\} a_{s}(\omega'+\omega) - a_{s}^\dag (\omega') \left[
  h_s(\omega'+\omega) -\ \gamma g^r_{\alpha\alpha}(\omega'+\omega)
  a_{s}(\omega'+\omega) \right] \right\rangle \\ & = &
\frac{i\gamma}{2} \int \frac{d\omega'}{2\pi} \left( \left[
  \left\langle h_s^\dag (\omega') a_{s}(\omega'+\omega) \right\rangle
  - \leg a_s^\dag(\omega') h_s(\omega'+\omega) \rg \right] - \gamma
\leg a_{s}^\dag(\omega') a_{s}(\omega'+\omega) \rg \left\{ \left[
  g_{\alpha\alpha}^{r}(\omega') \right]^* - g_{\alpha\alpha}^r
(\omega'+\omega) \right\} \right).
\label{eq:J_s_detail}
\end{eqnarray}
\end{widetext}
Employing Eq. (\ref{eq:c0_h}), we can derive the following relations:
\begin{equation}
\leg h_s^\dag(\omega')\, a_{s}(\omega) \rg = -\gamma\,
G_{ss}^r(\omega,\omega')\, I_\alpha(\omega'),
\end{equation}
\begin{equation}
\leg a_{s}^\dag (\omega')\, h_s(\omega'+\omega) \rg = -\gamma\,
G_{ss}^{r} (\omega',\omega'+\omega)\, I(\omega'+\omega),
\end{equation}
and
\begin{eqnarray}
\leg a_{s}^\dag(\omega')\, a_{s}(\omega) \rg & = & \gamma^2 \sum_{s'}
\int d\omega'' [G_{ss'}^r(\omega',\omega'')]^* G_{ss'}^r
(\omega,\omega'') \nn \\ & & \times\, I_\alpha(\omega'').
\end{eqnarray}
Putting these relations together with Eq. (\ref{eq:J_s_detail}) one
arrives at Eq. (\ref{eq:J_s}).


\begin{widetext}
  
\section{Spin current for 2D systems}
\label{sec:curr-expr-2d}

The fermionic particle operator in terms of the system Green's
function reads
\begin{align}
  \bs{c}_{js}^\dag (\omega) &= -\sum_{s';m,n} d\omega'
  h_{ns'}^\dag(\omega_1)\, \gamma_{n,m}^*
  \begin{bmatrix}
    G_{N,m,s';j,1,s}(\omega_1,\omega) & G_{N,m,s';j,2,s}(\omega_1,\omega) &
    \ldots & G_{N,m,s';j,d_j,s}(\omega_1,\omega),
    \end{bmatrix}
\end{align}
where $d_j$ is the number of sites in the slice $j$. After
substituting it into the current expression
\begin{equation}
J_{j}^z(\omega) = \frac{i}{2} \int \frac{d\omega'}{2\pi} \left[
  \bs{c}_{j}^\dag(\omega') \left( \sigma^z \otimes I_{d_{j}} \right)
  \bs{\tau}_{j-1}^\dag \,\bs{c}_{j-1}(\omega'+\omega) -
  \bs{c}_{j-1}^\dag(\omega')\, \bs{\tau}_{j-1} \left( \sigma^z \otimes
  I_{d_{j}} \right) \bs{c}_{j}(\omega'+\omega) \right],
\label{eq:sc2dj}
\end{equation}
the expectation value of the first term in Eq. (\ref{eq:sc2dj})
becomes
\begin{eqnarray}
  && \frac{i}{2} \int \frac{d\omega'}{2\pi} \sum_{s_1,s_2}\sum_{n,n';m,
    m'} \int d\omega_1 \gamma_{m',m}^*
  \begin{bmatrix}
    G_{N,m,s_1;j,1}^a(\omega_1,\omega) & G_{N,m,s_1;j,2}^a(\omega_1,\omega) &
    \ldots & G_{N,m,s_1;j,d_j}^a(\omega_1,\omega)
  \end{bmatrix}\nn\\
  && \times \left[ (\sigma^z\otimes I_{d_j}) \tau_{j-1}^\dag \right]
  \int d\omega_2
  \begin{bmatrix}
    G_{j-1,1;N,n,s_2}^r(\omega'+\omega,\omega_2) \\
    G_{j-1,2;N,n,s_2}^r(\omega'+\omega,\omega_2) \\
    \vdots \\
    G_{j-1,d_j;N,n,s_2}^r(\omega'+\omega,\omega_2)
  \end{bmatrix}
  \gamma_{n,n'}
  \leg h_{m',s_1}^\dag(\omega_1) h_{n',s_2}(\omega_2) \rg.
\end{eqnarray}
By applying the $h_{m,s}(\omega)$ correlator we find
\begin{eqnarray}
  \leg J_j^z(\omega \rg &=& \frac{i}{2} \int \frac{d\omega'}{2\pi}
  \int d\omega_1 {\rm Tr} \left[ \bs{\gamma}_\alpha^\dag
    \bs{G}_{N;j}^a(\omega_1,\omega) (\sigma^z\otimes \bs{I}_{d_j})
    \bs{\tau}_{j-1}^\dag \bs{G}_{j-1;N}^r(\omega+\omega',\omega_1)
    \bs{\gamma}_\alpha \bs{I}(\omega_1)\right]\nn\\ & &-\ \frac{i}{2}
  \int \frac{d\omega'}{2\pi} \int d\omega_1 {\rm Tr} \left[
    \bs{\gamma}_\alpha^\dag \bs{G}_{N;j-1}^a(\omega_1,\omega)
    \bs{\tau}_{j-1}(\sigma^z\otimes I_{d_j})
    \bs{G}_{j-1;N}^r(\omega+\omega',\omega_1) \bs{\gamma}_\alpha
    \bs{I}(\omega_1) \right].
\end{eqnarray}

We can follow the same approach to calculate the current at the
chain-reservoir interface:
\begin{equation}
  J_{\alpha}^z(\omega) = \frac{i}{2} \int
  \frac{d\omega'}{2\pi}  \left[ \bs{d}^\dag(\omega')(\sigma_z\otimes
    \bs{I}_L) \bs{\gamma}_\alpha
    \bs{c}_{N}(\omega'+\omega) - \bs{c}_{N}^\dag(\omega')\bs{\gamma}_\alpha^\dag
    (\sigma_z\otimes \bs{I}_L) \bs{d}_s(\omega'+\omega) \right].
\end{equation}
The current expression can be written as
\begin{eqnarray}
  J_{\alpha}^z(\omega) &=& \frac{i}{2} \int \frac{d\omega'}{2\pi}
  \left\{ \left[\bs{h}^\dag(\omega') - \bs{c}_{N}^\dag(\omega')
    \bs{\gamma}_\alpha^\dag \bs{g}_{\alpha\alpha}^a(\omega')\right]
  (\sigma_z\otimes \bs{I}_L) \bs{\gamma}_\alpha
  \bs{c}_{N}(\omega'+\omega) \right. \nn\\ &&
  \left. -\ \bs{c}_{N}^\dag(\omega')\bs{\gamma}_\alpha^\dag
  (\sigma_z\otimes \bs{I}_L)\left[ \bs{h}(\omega'+\omega) -
    \bs{g}_{\alpha\alpha}^r(\omega'+ \omega) \bs{\gamma}_\alpha
    \bs{c}_{N}(\omega+\omega') \right] \right\},
\end{eqnarray}
which can be simplified to
\begin{eqnarray}
  J_{\alpha}^z(\omega) &=& \frac{i}{2} \int \frac{d\omega'}{2\pi}
  \left[ \bs{h}^\dag(\omega')(\sigma_z\otimes
    \bs{I}_L) \bs{\gamma}_\alpha \bs{c}_{N}(\omega'+\omega) -
    \bs{c}_{N}^\dag(\omega') \bs{\gamma}_\alpha^\dag (\sigma_z\otimes
    \bs{I}_L) \bs{h}(\omega'+\omega) \right] \nn\\
  &&-\ \frac{i}{2} \int \frac{d\omega'}{2\pi}
  \bs{c}_{N}^\dag(\omega') \bs{\gamma}_\alpha^\dag \left[
    \bs{g}_{\alpha\alpha}^a(\omega') - \bs{g}_{\alpha\alpha}^r(\omega'+\omega) 
     \right](\sigma_z\otimes
    \bs{I}_L) \bs{\gamma}_\alpha \bs{c}_{N}(\omega+\omega').
\end{eqnarray}
After substituting the fermionic operator in terms of the system's
Green's function, the expectation value of the spin current becomes 
\begin{eqnarray}
  \leg J_{\alpha}^z(\omega) \rg &=& \frac{i}{2} \int
  \frac{d\omega'}{2\pi} \left\{  {\rm Tr}
  \left[ \bs{I}_\alpha(\omega'+\omega)\,\bs{\gamma}_\alpha 
    (\sigma_z\otimes
    \bs{I}_L) \bs{G}_{N;N}^a(\omega',\omega'+\omega) \bs{\gamma}_\alpha^\dag 
    \right]- {\rm Tr} \left[\bs{I}_\alpha(\omega')
    \bs{\gamma}_\alpha (\sigma_z\otimes
    \bs{I}_L) \bs{G}_{N;N}^r(\omega'+\omega,\omega')
    \bs{\gamma}_\alpha^\dag \right] \right\}  \nn\\
  &&+\ \frac{i}{2} \int
  \frac{d\omega'}{2\pi}  {\rm Tr} \left\{\bs{I}_\alpha(\omega'')\, \bs{\gamma}_\alpha
  \bs{G}_{N;N}^a(\omega',\omega'') \bs{\gamma}_\alpha^\dag\left[ 
     \bs{g}_{\alpha\alpha}^r(\omega'+\omega) - \bs{g}_{\alpha\alpha}^a(\omega')
     \right] (\sigma_z\otimes
    \bs{I}_L) \bs{\gamma}_\alpha
  \bs{G}_{N;N}^r(\omega'+\omega,\omega'')\bs{\gamma}_\alpha^\dag
  \right\} 
\end{eqnarray}
Similar to the 1D case, we expand the Green's function in terms of the
frequency difference,
\begin{equation}
  G(\omega,\omega') = \delta(\omega'-\omega)\, D_0(\omega) +
  \sum_{k\neq 0} \delta(\omega' - \omega - k\Omega)\, D_k(\omega),
\end{equation}
and following the same approach used in the 1D case, we get
\begin{eqnarray}
  \leg J_{\alpha}^z \rg &=& \frac{i}{2} \delta(\omega) \int
  \frac{d\omega'}{2\pi} {\rm Tr} \left\{\bs{I}_\alpha(\omega')\,
  \bs{\gamma}_\alpha (\sigma_z\otimes
    \bs{I}_L) \left[ \bs{D}_{0;N;N}^a(\omega') - \bs{D}_{0;N;N}^r(\omega')
    \right] \bs{\gamma}_\alpha^\dag \right\}\nn \\
  & &+\ \frac{i}{2} \delta(\omega) \int \frac{d\omega'}{2\pi}
  {\rm Tr}
  \left\{ \bs{I}_\alpha(\omega')\,  \bs{\gamma}_\alpha
  \bs{D}_{0;N;N}^a(\omega') \bs{\gamma}_\alpha^\dag 
  \left[ \bs{g}_{\alpha\alpha}^r(\omega') - \bs{g}_{\alpha\alpha}^a(\omega')
    \right] (\sigma_z\otimes\bs{I}_L)
  \bs{\gamma}_\alpha \bs{D}_{0;N;N}^r(\omega') \bs{\gamma}_\alpha^\dag
  \right\} \nn \\
  &&+\ \frac{i}{2} \delta(\omega) \int \frac{d\omega'}{2\pi}
  \sum_{k\neq 0} {\rm Tr} \left\{ \bs{I}_\alpha(\omega'-k\Omega)\,
  \bs{\gamma}_\alpha \bs{D}_{k;N;N}^a(\omega'-k\Omega) 
  \bs{\gamma}_\alpha^\dag 
  \left[ \bs{g}_{\alpha\alpha}^r(\omega') - \bs{g}_{\alpha\alpha}^a(\omega')
    \right] (\sigma_z\otimes\bs{I}_L) \bs{\gamma}_\alpha
  \bs{D}_{-k;N;N}^r(\omega') \bs{\gamma}_\alpha^\dag  \right\} \nn \\
\end{eqnarray}
which leads to
\begin{eqnarray}
  \leg J_{\alpha}^z \rg &=& -\frac{i}{2} \delta(\omega) \int
  \frac{d\omega'}{2\pi} \sum_{k\neq 0} {\rm Tr} \Big\{
  \bs{I}_\alpha(\omega')\, \bs{\gamma}_\alpha
  \bs{D}_{k;N;N}^r(\omega') \bs{\gamma}_\alpha^\dag \left[
    \bs{g}_{\alpha\alpha}^r(\omega'+k\Omega) -
    \bs{g}_{\alpha\alpha}^a(\omega' + k\Omega)
    \right](\sigma_z\otimes\bs{I}_L) \nn \\ & & \times\,
  \bs{\gamma}_\alpha \bs{D}_{-k;N;N}^a(\omega'+k\Omega)
  \bs{\gamma}_\alpha^\dag \Big\}\nn\\ &&+\ \frac{i}{2} \delta(\omega)
  \int \frac{d\omega'}{2\pi} \sum_{k\neq 0} {\rm Tr} \left\{
  \bs{I}_\alpha(\omega'+ k\Omega)\, \bs{\gamma}_\alpha
  \bs{D}_{k;N;N}^r(\omega') \bs{\gamma}_\alpha^\dag \left[
    \bs{g}_{\alpha\alpha}^r(\omega') -
    \bs{g}_{\alpha\alpha}^a(\omega') \right](\sigma_z\otimes\bs{I}_L)
  \bs{\gamma}_\alpha \bs{D}_{-k;N;N}^a(\omega'+k\Omega)
  \bs{\gamma}_\alpha^\dag \right\}, \nn \\
\end{eqnarray}
leading to
\begin{eqnarray}
  \overline{\leg J_{\alpha}^z(t) \rg} &=& \frac{1}{2} \int d\omega
  \sum_{k\neq 0}  \left[ f(\omega+k\Omega) - f(\omega) \right] \nn \\
  && \times {\rm Tr} \left[\bs{\rho}_\alpha(\omega+k\Omega)
    \bs{\gamma}_\alpha \bs{D}_{k;N;N}^r(\omega') 
    \bs{\gamma}_\alpha^\dag (\sigma^z\otimes \bs{I}_L)\,
    \bs{\rho}_\alpha(\omega)\, \bs{\gamma}_\alpha
    \bs{D}_{-k;N;N}^a(\omega'+ k\Omega) \bs{\gamma}_\alpha^\dag \right].
\end{eqnarray}
\end{widetext}




\begin{thebibliography}{27}
\bibitem{note} The spin current operator should actually be a tensor
  since, in addition to the direction of flow, the spin orientation
  determines the torque transfer direction. For a one-dimensional
  discrete chain, propagation can only happen along the chain and the
  spin current reduces to a vector.
\end{thebibliography}
\end{document}